\documentclass[a4paper,11pt]{article}

\usepackage{amsmath}
\usepackage{jheppub}

\usepackage{tikz}
\usetikzlibrary{calc,decorations.markings,arrows.meta}

\definecolor{c1}{RGB}{0, 114, 178}
\definecolor{c2}{RGB}{213, 94, 0}
\definecolor{c3}{RGB}{0, 158, 115}
\definecolor{c4}{RGB}{204, 121, 167}
\definecolor{c5}{RGB}{230, 159, 0}
\definecolor{c6}{RGB}{218, 165, 32}
\definecolor{myforestgreen}{RGB}{34, 139, 34}

\DeclareMathOperator{\crr}{cr}
\DeclareMathOperator{\li}{Li}
\DeclareMathOperator{\qli}{QLi}
\DeclareMathOperator{\Li}{Li}
\DeclareMathOperator{\QLi}{QLi}
\DeclareMathOperator{\Gr}{Gr}

\title{de Sitter Wavefunction from Quadrangular Polylogarithms: Chain Graphs}

\author[1]{Livia Ferro,}
\emailAdd{l.ferro@herts.ac.uk}

\author[1]{Tomasz {\L}ukowski,}
\emailAdd{t.lukowski@herts.ac.uk}

\author[2]{Lecheng Ren,}
\emailAdd{lecheng.ren@qmul.ac.uk}

\author[3,4]{Marcus Spradlin,}
\emailAdd{marcus\_spradlin@brown.edu}

\author[3]{Anastasia Volovich,}
\emailAdd{anastasia\_volovich@brown.edu}

\author[3]{He-Chen Weng}
\emailAdd{he-chen\_weng@brown.edu}

\author[1]{and Yao-Qi Zhang}
\emailAdd{y.zhang59@herts.ac.uk}

\affiliation[1]{Department of Physics, Astronomy and Mathematics,
University of Hertfordshire,
Hatfield, Hertfordshire, AL10 9AB, United Kingdom}

\affiliation[2]{Centre for Theoretical Physics, Department of Physics and Astronomy,
Queen Mary University of London,
London, E1 4NS, United Kingdom}

\affiliation[3]{Department of Physics,
Brown University,
Providence, RI 02912, USA}

\affiliation[4]{Brown Center for Theoretical Physics and Innovation,
Brown University,
Providence, RI 02912, USA}

\abstract{We present an explicit formula for the $n$-site chain graph contribution to the cosmological wavefunction for conformally coupled $\phi^3$ theory in de Sitter space. Our result relies on the recent finding that the symbol of this function satisfies total compatibility with respect to the $A_{2n-2}$ cluster algebra, and that Rudenko's quadrangular polylogarithms provide, by construction, a complete basis for such functions.  We prove our formula by directly relating a recursive set of differential equations satisfied by these wavefunction coefficients to a recursive coproduct formula for quadrangular polylogarithms.}

\begin{document}

\maketitle

\setlength{\fboxsep}{6pt}

\section{Introduction}

It is a recurring theme in theoretical physics that mathematical frameworks developed in a purely formal context often find physical realization only years or decades later. This perspective has been central to modern theoretical physics, particularly in our quest to understand scattering processes in particle physics and the evolution of the universe. However, there are many constructs that already exist in mathematics but do not yet have a clear application to physics. In this paper, we identify a physical realization for one such mathematical structure, and demonstrate that a class of functions recently defined in the mathematical literature, the quadrangular polylogarithms of Rudenko~\cite{rudenko2023goncharov, matveiakin2022cluster}, provides a natural language for describing the cosmological wavefunctions in de Sitter (dS) space~\cite{Arkani-Hamed:2017fdk}.

The study of scattering amplitudes in $\mathcal{N}=4$ supersymmetric Yang-Mills (SYM) theory has revolutionized our understanding of quantum field theory (see \cite{Arkani-Hamed:2022rwr} for reviews). A major ingredient in this progress is the discovery that the singularities of these amplitudes are governed by cluster algebras~\cite{Golden:2013xva}. The hypothesis that all symbol letters~\cite{Goncharov:2010jf} of six- and seven-point amplitudes in SYM theory are cluster variables of the $\Gr(4,6)$ and $\Gr(4,7)$ cluster algebras, and moreover satisfy a property known as cluster adjacency~\cite{Drummond:2017ssj,Drummond:2018dfd, Golden:2019kks}, has enabled an impressive bootstrap program which has led to the calculation of these amplitudes through five and eight loops, respectively~\cite{Dixon:2011pw,Dixon:2016nkn,Caron-Huot:2020bkp,Dixon:2023kop,He:2025tyv}. Cluster adjacency is the principle that two cluster variables appear as adjacent entries in the symbol of a multiple polylogarithm (MPL) function only if they belong to a common cluster. While certain patterns in adjacent triples have been observed~\cite{Drummond:2017ssj}, and the Landau equations impose general and powerful constraints on sequential symbol letters for individual Feynman diagrams~\cite{Hannesdottir:2022xki}, there is no known general long-range relation between cluster compatibility and symbol letters for SYM amplitudes.

The conceptual and computational machinery developed within the amplitudes program has proven broadly applicable to various corners of theoretical physics. This transfer of ideas has been particularly impactful in the study of curved geometries, including in dS and more generally FRW (see \cite{Benincasa:2022gtd, Baumann:2022jpr, Benincasa:2025uie} for review). The recent advances in cosmology have demonstrated that the wavefunction coefficients, much like their scattering amplitude counterparts, satisfy sophisticated differential equations~\cite{Arkani-Hamed:2023kig, De:2024zic, Capuano:2025ehm} and are governed by hidden mathematical symmetries.  Most strikingly, for particular contributions to the wavefunction coming from chain graphs in these cosmological backgrounds, their analytic structure is also dictated by ($A$-type) cluster algebras~\cite{Mazloumi:2025pmx,Capuano:2025myy}. (These algebras also appear in the multi-Regge limit of amplitudes in SYM theory~\cite{DelDuca:2016lad}.) This suggests that the rich mathematical structures discovered in the context of gauge theories are not incidental features of flat space, but are instead fundamental properties of quantum field theory that remain true across different gravitational backgrounds. 

While cluster adjacency was a key development in the context of SYM theory, the structures emerging in cosmology appear to be governed by even more restrictive and powerful constraints. Recent investigations into the cosmological wavefunctions have observed that not only does cluster adjacency hold true~\cite{Paranjape:2026htn}, but a stronger version of this property has emerged: total cluster compatibility~\cite{Capuano:2026pgq}. In cosmology, it is no longer just adjacent letters in the symbol that must be compatible, but rather the constraint becomes \emph{global}, requiring that every letter in a given word of the symbol reside within a single cluster. 
This observation is very significant. While the MPLs appearing in SYM theory amplitudes generally do not satisfy the total cluster compatibility, this property is the defining characteristic of the quadrangular polylogarithms introduced by Rudenko~\cite{rudenko2023goncharov, matveiakin2022cluster}, where they played a crucial role in proving Goncharov's depth conjecture. They are also the building blocks for alternating polylogarithms, which Rudenko used to provide an explicit formula for the volumes of hyperbolic orthoschemes, in terms of which the one-loop scalar $2n$-gon integral in $2n$-dimensions can be written~\cite{Ren:2023tuj}. While these remarkable quadrangular polylogarithm functions have not yet found a more direct application to amplitudes to date, they are precisely the correct building blocks for dS cosmological wavefunction coefficients.

In this paper, we provide the first explicit, closed-form expressions~\eqref{eq:generalP}, \eqref{eq:H_function_final} for the cosmological wavefunction coefficients associated with the complete class of chain graphs in the dS cosmology. We demonstrate that these contributions are naturally expressed as combinations of quadrangular polylogarithms. This result represents a significant simplification of the analytic structure of these observables, reducing complex nested integrals to a well-defined class of transcendental functions. To establish the validity of our expressions, we reformulate the existing differential and recursive relations satisfied by these wavefunctions~\cite{Hillman:2019wgh,He:2024olr}, and then compare them with the expressions for the coproduct of quadrangular polylogarithms derived by Rudenko. This allows us to prove that our result holds true for any $n$-site chain graph. To make these results accessible, we also include detailed examples of lower-weight cases, as well as a self-contained review of the ingredients required to define quadrangular polylogarithms.  

The paper is organized as follows: Section \ref{sec:prelude} reviews known results for two- and three-site chain graph wavefunction coefficients and identifies novel representations for them in terms of quadrangular polylogarithms, which we define in detail in Section \ref{sec:definitionofqli}. In Section~\ref{sec:recursion} we recall the recursion relations that wavefunction coefficient satisfies and rewrite them in the form that is more suitable for comparison with the definitions from mathematics. Then in Section \ref{sec:solution} we provide the solution to these recursion relations and write explicit expressions for wavefunction coefficients in terms of quadrangular polylogarithms. We also include the detailed proof of our formulas in the appendix. An ancillary file contains machine-readable formulas for the wavefunction coefficients associated to two-, three-, and four-site chain graphs. 

\section{Prelude: from dS Wavefunctions to Quadrangular Polylogarithms}
\label{sec:prelude}

Before describing our general solution for any chain, we begin by revisiting the wavefunction coefficients for the two- and three-site chains (the ``one-site chain" is a special case; see Section~\ref{sec:examples}).  By exploiting the fact that their symbols satisfy total cluster compatibility~\cite{Capuano:2026pgq}, we are led to new, compact representations for these functions which naturally suggest a connection to the quadrangular polylogarithms of Rudenko~\cite{rudenko2023goncharov, matveiakin2022cluster}.

\subsection{Two-site Chain}
\label{sec:prelude2}

The symbol of the wavefunction coefficient for the two-site chain
\begin{align}
    \psi_{2}(X_1,X_2; Y_{1,2}) =
\begin{tikzpicture}[baseline=(current bounding box.center),scale=1.3]
		\draw[line width=1.5pt] (0,0)--(1.25,0);
		\draw[fill=black] (0,0) circle [radius=0.075];
		\draw[fill=black] (1.5-0.2,0) circle [radius=0.075];
		\node[below] at (0,-0.1) {{$X_1$}};
		\node[below] at (1.5-0.25,-0.1) {{$X_{2}$}};
		\node[above] at (0.625,0.01) {{${Y}_{1,2}$}};
\end{tikzpicture}
\label{eq:psi2}
\end{align}
is \cite{Arkani-Hamed:2017fdk,Hillman:2019wgh, Arkani-Hamed:2023kig}
\begin{equation}\label{eq:P2symbolXY}
\mathcal{S}(\psi_{2})= \frac{X_1+Y_{1,2}}{X_1+X_2}\otimes \frac{X_2-Y_{1,2}}{X_2+Y_{1,2}} + \frac{X_2+Y_{1,2}}{X_1+X_2}\otimes \frac{X_1-Y_{1,2}}{X_1+Y_{1,2}}\,.
\end{equation}
We can embed these variables into the Grassmannian $G(2,5)$ via the $2 \times 5$ matrix representative~\cite{Capuano:2025myy,Mazloumi:2025pmx,Paranjape:2026htn,Capuano:2026pgq}
\begin{equation}
C_{2}=    \left(
\begin{array}{ccccc}
 1 & 1 & 1 & 1 & 0 \\
 0 &X_1+Y_{1,2} & X_1-Y_{1,2} & X_1+X_2 & 1 \\
\end{array}
\right).
\end{equation}
Then, in terms of Pl\"ucker variables $\Delta_{ij}$, \eqref{eq:P2symbolXY} reads
\begin{equation}
    \mathcal{S}(\psi_{2})=\frac{\Delta_{12}}{\Delta_{14}}\otimes \frac{\Delta_{24}}{\Delta_{34}}+\frac{\Delta_{34}}{\Delta_{14}}\otimes \frac{\Delta_{13}}{\Delta_{12}}\,.
\end{equation}
This answer is not invariant under rescaling of columns of the matrix $C_{2}$, because it is expressed in a gauge where $\Delta_{i5} = 1$ for all $i$. There is a unique way to un-gauge fix by uplifting each term to cross-ratios using
\begin{equation}
    \frac{\Delta_{ij}}{\Delta_{ik}} \rightarrow \frac{\Delta_{ij}\Delta_{k5}}{\Delta_{ik}\Delta_{j5}}\,,
\end{equation}
which leads to the projectively invariant symbol
\begin{align}
\label{eq:2chain1}
  \mathcal{S}(\psi_{2})=\frac{\Delta_{12}\Delta_{45}}{\Delta_{14}\Delta_{25}}\otimes   \frac{\Delta_{24}\Delta_{35}}{\Delta_{34}\Delta_{25}}+\frac{\Delta_{34}\Delta_{15}}{\Delta_{14}\Delta_{35}}\otimes   \frac{\Delta_{13}\Delta_{25}}{\Delta_{12}\Delta_{35}}
\end{align}
that is well-defined on $\mathcal{M}_{0,5}$. This formula does not manifestly satisfy cluster compatibility, which is the condition that when expanded out, no $\Delta_{ij}$ can appear next to $\Delta_{kl}$ if $ij$ and $kl$ cross each other (when thought of as chords in a pentagon). For example, expanding out only the first term gives a contribution $+\Delta_{14} \otimes \Delta_{25}$ that involves two incompatible Pl\"ucker coordinates. However, the symbol can be rearranged into the form
\begin{equation}
\label{eq:2chain2}
   \mathcal{S}(\psi_{2}) = \frac{\Delta_{23}\Delta_{15}}{\Delta_{12}\Delta_{35}}\otimes \frac{\Delta_{13}\Delta_{25}}{\Delta_{12}\Delta_{35}}-\frac{\Delta_{14}\Delta_{23}}{\Delta_{34}\Delta_{12}}\otimes \frac{\Delta_{13}\Delta_{24}}{\Delta_{34}\Delta_{12}} +\frac{\Delta_{23}\Delta_{45}}{\Delta_{25}\Delta_{34}}\otimes \frac{\Delta_{24}\Delta_{35}}{\Delta_{25}\Delta_{34}}\,,
\end{equation}
where cluster compatibility is now manifest term-by-term.
However, in order to write the manifestly cluster compatible expression~(\ref{eq:2chain2}) it is necessary to introduce the spurious symbol letter $\Delta_{23} = -2 Y_{1,2}$ which is absent from~(\ref{eq:2chain1}). If we define the cross-ratios
\begin{equation}
\label{eq:qdef}
    q_{ijkl}=\frac{\Delta_{ij}\Delta_{kl}}{\Delta_{jk}\Delta_{li}}\,,
\end{equation}
we can further rewrite~(\ref{eq:2chain2}) as
\begin{equation}
\label{eq:2chain3}
    \mathcal{S}(\psi_{2}) = q_{5123}\otimes (1-q_{5123})-q_{4123}\otimes(1-q_{4123}) +q_{4523}\otimes(1-q_{4523})\,.
\end{equation}
It is now trivial to integrate~(\ref{eq:2chain3}) to arrive at
\begin{equation}
\label{eq:2chain4}
   \psi_{2} = -\li_2(1-q_{5123})+\li_2(1-q_{4123})-\li_2(1-q_{4523})+\mathcal{O}(\pi)\,,
\end{equation}
and terms not captured by the symbol can be fixed by imposing that $\psi_2$ should vanish in the soft limit $Y_{1,2} \to 0$. In this manner we arrive at the final answer, which we choose to write as
\begin{equation}\label{eq:P2H}
\boxed{
\psi_{2} = F_2(5,1,2,3)-F_2(4,1,2,3)+F_2(4,5,2,3)
}
\end{equation}
in terms of the function
\begin{equation}
\label{eq:Fquad}
\boxed{
    F_2(a,b,c,d) = \begin{tikzpicture}[baseline=(current bounding box.center), scale=0.7]

\foreach \i in {0,...,3} {
    \coordinate (V\i) at ({cos(90 - \i*90 + 45)}, {sin(90 - \i*90 + 45)});
}

\node at ({1.25*cos(90 - 0*90 + 45)}, {1.25*sin(90 - 0*90 + 45)}) {$b$};
\node at ({1.25*cos(90 - 1*90 + 45)}, {1.25*sin(90 - 1*90 + 45)}) {$c$};
\node at ({1.25*cos(90 - 2*90 + 45)}, {1.25*sin(90 - 2*90 + 45)}) {$d$};
\node at ({1.25*cos(90 - 3*90 + 45)}, {1.25*sin(90 - 3*90 + 45)}) {$a$};
\draw (V0) -- (V1);
\draw (V1) -- (V2);
\draw (V2) -- (V3) node[midway] {$\times$};
\draw (V3) -- (V0);
\end{tikzpicture} = \Li_2(q_{abcd})-\Li_1(q_{abcd})\log(q_{abcd})}
\end{equation}
associated to a rooted quadrilateral. The marked edge between $a$ and $d$, which we call the root, is a reminder that $F_2$ is not cyclically invariant in its arguments.  It does however satisfy $F_2(a,b,c,d) = F_2(d,c,b,a)$ so we do not need to orient the quadrilateral.

One can check that the formula~(\ref{eq:P2H}) agrees with the known answer, given in equation (5.5) of~\cite{Hillman:2019wgh}, but the former has manifest geometric structure that we will see generalizes nicely to arbitrary chains.  It is also interesting to note that the representation~(\ref{eq:P2H}) has no ``beyond the symbol'' terms.

\subsection{Three-site Chain}
\label{sec:prelude3}

We can perform similar steps for the symbol of the three-site chain wavefunction coefficient
\begin{align}
    \psi_{3}(X_1,X_2,X_3; Y_{1,2},Y_{2,3}) =
\begin{tikzpicture}[baseline=(current bounding box.center),scale=1.3]
		\draw[line width=1.5pt] (0,0)--(2.75,0);
		\draw[fill=black] (0,0) circle [radius=0.075];
		\draw[fill=black] (1.5-0.2,0) circle [radius=0.075];
		\draw[fill=black] (2.5+0.2,0) circle [radius=0.075];
		\node[below] at (0,-0.1) {{$X_1$}};
		\node[below] at (1.5-0.25,-0.1) {{$X_{2}$}};
		\node[below] at (2.5+0.25,-0.1) {{$X_{3}$}};
		\node[above] at (0.625,0.01) {{${Y}_{1,2}$}};
        \node[above] at (2,0.01) {{$Y_{2,3}$}};
\end{tikzpicture}
\label{eq:psi3}
\end{align}
which has been explicitly written in
\cite{Arkani-Hamed:2017fdk,Hillman:2019wgh, Arkani-Hamed:2023kig}
and contains 104 terms. We embed the variables into $G(2,7)$ via
\begin{equation}
 C_{3} =  \left(
\begin{array}{ccccccc}
 1 & 1 & 1 & 1 & 1 & 1 & 0 \\
 0 & X_1+Y_{1,2} & X_1-Y_{1,2} & X_1+X_2+Y_{2,3} & X_1+X_2-Y_{2,3} & X_1+X_2+X_3 & 1 \\
\end{array}
\right),
\end{equation}
express the symbol in terms of Pl\"ucker variables, and then uplift them to cross-ratios as before. We find that it can then be expressed as
\begin{equation}\label{eq:P3H}
    \mathcal{S}(\psi_{3})=\mathcal{S}(F_3(7,1,2,3,4,5))-\mathcal{S}(F_3(6,1,2,3,4,5))+\mathcal{S}(F_3(6,7,2,3,4,5))\,,
\end{equation}
where
\begin{align}
\label{eq:F3}
   \mathcal{S}( F_3(a,b,c,d,e,f))&=q_{abcf}\otimes \left(1-q_{abcf}\right)\otimes \left(1-q_{cdef}\right)-q_{abcd}\otimes \left(1-q_{abcd}\right)\otimes \left(1-q_{adef}\right) \nonumber\\
    &-q_{abef}\otimes \left(1-q_{abef}\right)\otimes \left(1-q_{cdeb}\right) +q_{adef}\otimes \left(1-q_{adef}\right)\otimes \left(1-q_{abcd}\right)\nonumber\\
    &-q_{cdef}\otimes \left(1-q_{cdef}\right)\otimes \left(1-q_{abcf}\right)+q_{cdeb}\otimes \left(1-q_{cdeb}\right)\otimes \left(1-q_{abef}\right) \,.
\end{align}

Like in the two-site case~\eqref{eq:2chain2}, each term in~\eqref{eq:P3H} contains both of the spurious symbol letters $\Delta_{23} = -2 Y_{1,2}$ and $\Delta_{45} = -2 Y_{2,3}$, which cancel out in the sum. Let us now explain how the representation~(\ref{eq:P3H}) was found.  Guided by the crucial observation that the symbol satisfies total cluster compatibility~\cite{Capuano:2026pgq}, one starts with an ansatz that consists of compatible words, where two cross-ratios are compatible if they correspond to non-intersecting quadrangles in a hexagon. If $a,b,c,d$ are in cyclic order then any cross-ratio $q_{abcd}$ is compatible with itself and with $(1-q_{abcd})$, while $(1-q_{abcd})$ is not compatible with itself. Consequently, any letter of the form $(1-q_{abcd})$ can appear at most once in any valid word. We find that~(\ref{eq:P3H}) is the unique linear combination of valid words that matches the symbol of $\psi_3$, and note the interesting feature that the coefficient of every allowed term is $\pm 1$. The symbol~(\ref{eq:F3}) can be integrated to a function and finally we obtain
\begin{align}
\label{eq:3chainfinal}
\boxed{
    \psi_{3} = F_3(7,1,2,3,4,5) - F_3(6,1,2,3,4,5) + F_3(6,7,2,3,4,5)\,,}
\end{align}
where $F_3(a,b,c,d,e,f)$ is a sum of four terms naturally associated to a rooted hexagon and its quadrangulations:
\begin{equation}
\label{eq:hexagonF}
\boxed{
F_3(a,b,c,d,e,f) = \left\{
\begin{aligned}
&\begin{tikzpicture}[baseline=(current bounding box.center), scale=0.7]

\foreach \i in {0,...,6} {
    \coordinate (V\i) at ({cos(90 - \i*60 + 30)}, {sin(90 - \i*60 + 30)});
}

\node at ({1.35*cos(90 - 0*60 + 30)}, {1.35*sin(90 - 0*60 + 30)}) {$c$};
\node at ({1.35*cos(90 - 1*60 + 30)}, {1.35*sin(90 - 1*60 + 30)}) {$d$};
\node at ({1.35*cos(90 - 2*60 + 30)}, {1.35*sin(90 - 2*60 + 30)}) {$e$};
\node at ({1.35*cos(90 - 3*60 + 30)}, {1.35*sin(90 - 3*60 + 30)}) {$f$};
\node at ({1.35*cos(90 - 4*60 + 30)}, {1.35*sin(90 - 4*60 + 30)}) {$a$};
\node at ({1.25*cos(90 - 5*60 + 30)}, {1.35*sin(90 - 5*60 + 30)}) {$b$};
\draw (V0) -- (V1);
\draw (V1) -- (V2);
\draw (V2) -- (V3);
\draw (V5) -- (V0);
\draw (V4) -- (V5);
\draw (V3) -- (V4)  node[midway] {$\times$};
\end{tikzpicture} =  \begin{aligned}&-(\QLi^+_3(a,b,c,d,e,f)\\&+\QLi^+_2(a,b,c,d,e,f)\log(q_{abcdef}))\end{aligned} \\
&\quad\quad~\,+\\
&\begin{tikzpicture}[baseline=(current bounding box.center), scale=0.7]

\foreach \i in {0,...,6} {
    \coordinate (V\i) at ({cos(90 - \i*60 + 30)}, {sin(90 - \i*60 + 30)});
}

\node at ({1.35*cos(90 - 0*60 + 30)}, {1.35*sin(90 - 0*60 + 30)}) {$c$};
\node at ({1.35*cos(90 - 1*60 + 30)}, {1.35*sin(90 - 1*60 + 30)}) {$d$};
\node at ({1.35*cos(90 - 2*60 + 30)}, {1.35*sin(90 - 2*60 + 30)}) {$e$};
\node at ({1.35*cos(90 - 3*60 + 30)}, {1.35*sin(90 - 3*60 + 30)}) {$f$};
\node at ({1.35*cos(90 - 4*60 + 30)}, {1.35*sin(90 - 4*60 + 30)}) {$a$};
\node at ({1.25*cos(90 - 5*60 + 30)}, {1.35*sin(90 - 5*60 + 30)}) {$b$};
\draw (V0) -- (V3);
\draw (V0) -- (V1);
\draw (V1) -- (V2);
\draw (V2) -- (V3);
\draw (V5) -- (V0);
\draw (V4) -- (V5);
\draw (V3) -- (V4)  node[midway] {$\times$};
\end{tikzpicture} =  -F_2(a,b,c,f)\Li_1(q_{cdef}) \\
&\quad\quad~\,+\\
&\begin{tikzpicture}[baseline=(current bounding box.center), scale=0.7]

\foreach \i in {0,...,6} {
    \coordinate (V\i) at ({cos(90 - \i*60 + 30)}, {sin(90 - \i*60 + 30)});
}

\node at ({1.35*cos(90 - 0*60 + 30)}, {1.35*sin(90 - 0*60 + 30)}) {$c$};
\node at ({1.35*cos(90 - 1*60 + 30)}, {1.35*sin(90 - 1*60 + 30)}) {$d$};
\node at ({1.35*cos(90 - 2*60 + 30)}, {1.35*sin(90 - 2*60 + 30)}) {$e$};
\node at ({1.35*cos(90 - 3*60 + 30)}, {1.35*sin(90 - 3*60 + 30)}) {$f$};
\node at ({1.35*cos(90 - 4*60 + 30)}, {1.35*sin(90 - 4*60 + 30)}) {$a$};
\node at ({1.25*cos(90 - 5*60 + 30)}, {1.35*sin(90 - 5*60 + 30)}) {$b$};
\draw (V1) -- (V4);
\draw (V0) -- (V1);
\draw (V1) -- (V2);
\draw (V2) -- (V3);
\draw (V5) -- (V0);
\draw (V4) -- (V5);
\draw (V3) -- (V4)  node[midway] {$\times$};
\end{tikzpicture} = -F_2(a,d,e,f)\Li_1(q_{abcd}) \\
&\quad\quad~\,+\\
&\begin{tikzpicture}[baseline=(current bounding box.center), scale=0.7]

\foreach \i in {0,...,6} {
    \coordinate (V\i) at ({cos(90 - \i*60 + 30)}, {sin(90 - \i*60 + 30)});
}

\node at ({1.35*cos(90 - 0*60 + 30)}, {1.35*sin(90 - 0*60 + 30)}) {$c$};
\node at ({1.35*cos(90 - 1*60 + 30)}, {1.35*sin(90 - 1*60 + 30)}) {$d$};
\node at ({1.35*cos(90 - 2*60 + 30)}, {1.35*sin(90 - 2*60 + 30)}) {$e$};
\node at ({1.35*cos(90 - 3*60 + 30)}, {1.35*sin(90 - 3*60 + 30)}) {$f$};
\node at ({1.35*cos(90 - 4*60 + 30)}, {1.35*sin(90 - 4*60 + 30)}) {$a$};
\node at ({1.25*cos(90 - 5*60 + 30)}, {1.35*sin(90 - 5*60 + 30)}) {$b$};
\draw (V2) -- (V5);
\draw (V0) -- (V1);
\draw (V1) -- (V2);
\draw (V2) -- (V3);
\draw (V5) -- (V0);
\draw (V4) -- (V5);
\draw (V3) -- (V4)  node[midway] {$\times$};
\end{tikzpicture} =  + F_2(a,b,e,f)\Li_1(1/q_{bcde})\\
\end{aligned}\right.}
\end{equation}
where $F_2(a,b,c,d)$ is the quadrilateral function defined in~(\ref{eq:Fquad}) and $\QLi$ denotes the quadrangular polylogarithm function to be introduced in the next section. We also introduced a shorthand notation $q_{abcdef} = q_{abcd}q_{adef}$. Although the representation~(\ref{eq:3chainfinal}) looks very different than the one originally given in equation~(5.8) of~\cite{Hillman:2019wgh}, one can check numerically that they are equal. It is again interesting to note that~(\ref{eq:3chainfinal}) does not have any ``beyond the symbol'' terms, and exhibits clear geometric structure hinting that it should generalize.

\section{Mathematics: Rudenko's Quadrangular Polylogarithms}\label{sec:definitionofqli}

In this section we define, for $m \ge n \ge 1$, the even and odd \emph{quadrangular polylogarithm} functions $\QLi^\pm_m(0,\ldots,2n{+}1)$. These are weight-$m$ multiple polylogarithm functions of cross-ratios of points $z_0,\ldots,z_{2n+1} \in \mathbb{P}^1$ whose symbols satisfy total compatibility with respect to the $A_{2n-2}$ cluster algebra.  Our definition follows Theorem~1.2 of~\cite{rudenko2023goncharov}, but in that reference ``QLi'' is not used to denote specific functions but rather equivalence classes of functions modulo products of functions of lower weight.  In contrast we shall choose specific functional representatives that turn out to be surprisingly well-suited to the cosmological correlator problem.

\subsection{Arborification}

The first step in Rudenko's construction is to associate to every alternating polygon $P$ a certain weighted symbol $T_P$, which is an element of the Connes-Kreimer Hopf algebra of rooted trees~\cite{Connes:1998qv}. A \emph{weighted symbol} is like the ordinary symbol of multiple polylogarithm functions familiar to the physics literature from~\cite{Goncharov:2010jf}, but with each symbol letter $\varphi$ carrying an additional non-negative integer \emph{weight} $m$, denoted $[\varphi,m]$. An \emph{alternating $(2n{+}2)$-gon} is an ordered list of $2n{+}2$ integers $P = (p_0, p_1, \ldots, p_{2n+1})$ that alternate between even and odd.  $P$ is called even (odd) if the smallest entry $p_0$ is even (odd).  We can define $T_P$ recursively via the following rules:
\begin{enumerate}
\item Draw a convex $(2n{+}2)$-gon and label the corners $p_0,\ldots,p_{2n+1}$ in cyclic order.
\item Draw some quadrangulation $Q$ of $P$ and its dual graph $\widetilde{Q}$.
\item Color each node $\widetilde{Q}$ white (black) if the quadrangle it sits in is even (odd).
\item Mark the root node with a square; this is the one that contains the edge $(p_{2n+1},p_0)$.
\item Draw an arrow on each edge of $\widetilde{Q}$ pointing away from the root node.  This is unambiguous since $\widetilde{Q}$ is a tree.
\item Label each node $Q_i$ of $\widetilde{Q}$ by the weighted symbol letter $[\crr(Q_i),1]$ and the cross-ratio of an alternating quadrangle $(a,b,c,d)$ is defined in terms of~(\ref{eq:qdef}) by
\begin{align}
\crr(a,b,c,d) = \begin{cases}
q_{abcd} & a~{\rm even}\,,
\\
1/q_{abcd} & a~{\rm odd}\,.
\end{cases}
\label{eq:crdef}
\end{align}
\item Define a weighted symbol $t_Q$ associated to the quadrangulation recursively, starting from the root node $Q_0$, by the rule
\begin{align}
\label{eq:recursioneven}
\raisebox{-5pt}{\begin{tikzpicture}[baseline=(current bounding box.center), scale=1.5]
\draw ($(-0.43301,-0)+(-3pt,-3pt)$) rectangle ($(-0.43301,0)+(3pt,3pt)$);
\draw[postaction={decorate},decoration={markings,mark=at position 0.7 with {\arrow[scale=1.5]{Stealth}}},thick] (-0.433013,0) -- (-1.29904,0);
\draw[postaction={decorate},decoration={markings,mark=at position 0.7 with {\arrow[scale=1.5]{Stealth}}},thick] (-0.433013,0) -- (-0.433013,0.86626);
\draw[postaction={decorate},decoration={markings,mark=at position 0.7 with {\arrow[scale=1.5]{Stealth}}},thick] (-0.433013,0) -- (0.433013,0);
\filldraw[fill=white] (-0.433013,0) circle (2pt) node[below=6pt] {$[\crr(Q_0),1]$};
\filldraw[fill=gray] (0.433013,0) circle (2pt) node[right=4pt] {$t_3$};
\filldraw[fill=gray] (-1.29904,0) circle (2pt) node[left=4pt] {$t_1$};
\filldraw[fill=gray] (-0.433013,0.86626) circle (2pt) node[above=2pt] {$t_2$};
\end{tikzpicture}}
&= - [\crr(Q_0), 1] \otimes (t_1 \star t_2 \star t_3)\,,
\\
\raisebox{-5pt}{\begin{tikzpicture}[baseline=(current bounding box.center), scale=1.5]
\draw ($(-0.43301,-0)+(-3pt,-3pt)$) rectangle ($(-0.43301,0)+(3pt,3pt)$);
\draw[postaction={decorate},decoration={markings,mark=at position 0.7 with {\arrow[scale=1.5]{Stealth}}},thick] (-0.433013,0) -- (-1.29904,0);
\draw[postaction={decorate},decoration={markings,mark=at position 0.7 with {\arrow[scale=1.5]{Stealth}}},thick] (-0.433013,0) -- (0.433013,0);
\draw[postaction={decorate},decoration={markings,mark=at position 0.7 with {\arrow[scale=1.5]{Stealth}}},thick] (-0.433013,0) -- (-0.433013,0.86626);
\filldraw[fill=black] (-0.433013,0) circle (2pt) node[below=6pt] {$[\crr(Q_0),1]$};
\filldraw[fill=gray] (0.433013,0) circle (2pt) node[right=4pt] {$t_3$};
\filldraw[fill=gray] (-1.29904,0) circle (2pt) node[left=4pt] {$t_1$};
\filldraw[fill=gray] (-0.433013,0.86626) circle (2pt) node[above=2pt] {$t_2$};
\end{tikzpicture}}
&=+ [\crr(Q_0), 1] \otimes (t_1 \star t_2 \star t_3)
+ [\crr(Q_0), 1]  \cdot  (t_1 \star t_2 \star t_3)\,,
\label{eq:recursionodd}
\end{align}
where $t_i$ is the weighted symbol of the subtree rooted at $i$. If any of the subtrees $t_i$ are empty, the corresponding $t_i$ should be set to the empty word, which we denote by ``$1$", in the expressions on the right-hand side.
\item Finally, $T_P$ is given by the sum over all quadrangulations $\mathcal{Q}(P)$ of $P$:
\begin{align}
T_P = \sum_{Q \in \mathcal{Q}(P)} t_Q\,.
\end{align}
\end{enumerate}
Step~7 relies on the product defined by

\begin{align}
\label{eq:productdef}
[\varphi, m]  \cdot  \Big([\varphi_1,m_1 ] \otimes[\varphi_2,m_2] \otimes \cdots \Big) =
[\varphi\, \varphi_1, m{+}m_1] \otimes  [ \varphi_2,m_2 ] \otimes \cdots
\end{align}
and the star (or quasi-shuffle) product defined recursively by
\begin{multline}
    ([ \varphi_1, m_1 ] \otimes \omega_1) \star ([ \varphi_2, m_2 ] \otimes \omega_2)
    = [ \varphi_1, m_1 ] \otimes ( (\omega_1) \star ([ \varphi_2, m_2 ] \otimes \omega_2) )\\ + [ \varphi_2, m_2 ] \otimes ( ([ \varphi_1, m_1 ] \otimes \omega_1) \star (\omega_2) ) + ([ \varphi_1, m_1 ] \cdot [ \varphi_2, m_2 ]) \otimes (\omega_1 \star \omega_2)\,,
\label{eq:stardef}
\end{multline}
where $\omega_1$ and $\omega_2$ denote words of arbitrary length. The empty word 1 is the identity element under the star product, and the zero element under the dot product:
\begin{align}
\label{eq:identity}
    1 \star \omega = \omega \star 1 = \omega\,, \qquad [\varphi, m] \cdot 1 = 0\,.
\end{align}
The above definitions extend to linear combinations of words in the obvious way.

\subsubsection{Examples}

The base case for the recursive construction described in the previous section is when $P$ is a quadrilateral.
For the even quadrilateral $P=(0,1,2,3)$ we have from~(\ref{eq:crdef}) and (\ref{eq:recursioneven}) that
\begin{equation}
\label{eq:evenquad}
T_{(0,1,2,3)} =
\begin{tikzpicture}[baseline=(current bounding box.center), scale=1.0, rotate=90]

\foreach \i in {0,...,3} {
    \coordinate (V\i) at ({cos(90 - \i*90 + 45)}, {sin(90 - \i*90 + 45)});
}

\foreach \i in {0,...,3} {
    \pgfmathtruncatemacro{\j}{mod(\i+1,4)}
    \draw (V\i) -- (V\j);
}

\foreach \i in {0,...,3} {
    \node at ({1.25*cos(90 - \i*90 + 45)}, {1.25*sin(90 - \i*90 + 45)}) {\(\i\)};
}

\draw (-3pt,-3pt) rectangle (+3pt,+3pt);
\filldraw[fill=white] (0,0) circle (2pt);

\draw (V0) -- (V3)  node[midway] {$\times$};

\end{tikzpicture}
= - [q_{0123}, 1]\,,
\end{equation}
and for the odd quadrilateral $P=(1,2,3,4)$ we have from~(\ref{eq:crdef}), (\ref{eq:recursionodd}) and~(\ref{eq:identity}) that
\begin{equation}
T_{(1,2,3,4)}=
\begin{tikzpicture}[baseline=(current bounding box.center), scale=1.0, rotate=180]

\foreach \i in {0,...,3} {
    \coordinate (V\i) at ({cos(90 - \i*90 + 45)}, {sin(90 - \i*90 + 45)});
}

\foreach \i in {0,...,3} {
    \pgfmathtruncatemacro{\j}{mod(\i+1,4)}
    \draw (V\i) -- (V\j);
}

\foreach \i in {1,...,4} {
    \node at ({1.25*cos(90 - \i*90 + 45)}, {1.25*sin(90 - \i*90 + 45)}) {\(\i\)};
}

\draw (-3pt,-3pt) rectangle (+3pt,+3pt);
\fill (0,0) circle (2pt);

\draw (V0) -- (V1)  node[midway] {$\times$};

\end{tikzpicture}
= + [1/q_{1234}, 1]\,.
\label{eq:oddquad}
\end{equation}
Conventionally it is true that $[1/\varphi] = - [\varphi]$ at symbol level, which would make it seem that~(\ref{eq:oddquad}) is a redundant rewriting of~(\ref{eq:evenquad}), but it is important that we treat them as distinct formal symbols, since we will upgrade these to particular functions in the next section.

Next we consider hexagons, which have three quadrangulations.
First, for the even hexagon $P=(0,1,2,3,4,5)$ we have
\begin{equation}
\begin{tikzpicture}[baseline=(current bounding box.center), scale=1.0, rotate=120]

\foreach \i in {0,...,5} {
    \coordinate (V\i) at ({cos(90 - \i*60 + 30)}, {sin(90 - \i*60 + 30)});
}

\foreach \i in {0,...,5} {
    \pgfmathtruncatemacro{\j}{mod(\i+1,6)}
    \draw (V\i) -- (V\j);
}

\foreach \i in {0,...,5} {
    \node at ({1.25*cos(90 - \i*60 + 30)}, {1.25*sin(90 - \i*60 + 30)}) {\(\i\)};
}

\coordinate (C1) at (0,0.433013);
\coordinate (C2) at (0,-0.433013);

\draw[rotate=-120] ($(C1)+(-3pt,-3pt)$) rectangle ($(C1)+(3pt,3pt)$);

\draw[postaction={decorate},decoration={markings,mark=at position 0.7 with {\arrow[scale=1.5]{Stealth}}},thick] (C1) -- (C2);

\filldraw[fill=white] (C1) circle (2pt);
\filldraw[fill=white] (C2) circle (2pt);

\draw (V2) -- (V5);

\draw (V0) -- (V5)  node[midway] {$\times$};

\end{tikzpicture}
= + [q_{0125},1] \otimes [q_{2345},1]\,,
\end{equation}

\begin{equation}
\begin{tikzpicture}[baseline=(current bounding box.center), scale=1.0, rotate=120]

\foreach \i in {0,...,5} {
    \coordinate (V\i) at ({cos(90 - \i*60 + 30)}, {sin(90 - \i*60 + 30)});
}

\foreach \i in {0,...,5} {
    \pgfmathtruncatemacro{\j}{mod(\i+1,6)}
    \draw (V\i) -- (V\j);
}

\foreach \i in {0,...,5} {
    \node at ({1.25*cos(90 - \i*60 + 30)}, {1.25*sin(90 - \i*60 + 30)}) {\(\i\)};
}

\coordinate (C1) at (-0.375,0.216506);
\coordinate (C2) at (0.375,-0.216506);

\draw[rotate=-120] ($(C1)+(-3pt,-3pt)$) rectangle ($(C1)+(3pt,3pt)$);

\draw[postaction={decorate},decoration={markings,mark=at position 0.7 with {\arrow[scale=1.5]{Stealth}}},thick] (C1) -- (C2);

\filldraw[fill=white] (C1) circle (2pt);
\fill (C2) circle (2pt);

\draw (V1) -- (V4);

\draw (V0) -- (V5)  node[midway] {$\times$};

\end{tikzpicture}
= - [q_{0145},1] \otimes [1/q_{1234},1]\,,
\end{equation}

\begin{equation}
\begin{tikzpicture}[baseline=(current bounding box.center), scale=1.0, rotate=120]

\foreach \i in {0,...,5} {
    \coordinate (V\i) at ({cos(90 - \i*60 + 30)}, {sin(90 - \i*60 + 30)});
}

\foreach \i in {0,...,5} {
    \pgfmathtruncatemacro{\j}{mod(\i+1,6)}
    \draw (V\i) -- (V\j);
}

\foreach \i in {0,...,5} {
    \node at ({1.25*cos(90 - \i*60 + 30)}, {1.25*sin(90 - \i*60 + 30)}) {\(\i\)};
}

\coordinate (C1) at (0.375,0.216506);
\coordinate (C2) at (-0.375,-0.216506);

\draw[rotate=-120] ($(C2)+(-3pt,-3pt)$) rectangle ($(C2)+(3pt,3pt)$);

\draw[postaction={decorate},decoration={markings,mark=at position 0.7 with {\arrow[scale=1.5]{Stealth}}},thick] (C2) -- (C1);

\filldraw[fill=white] (C1) circle (2pt);
\filldraw[fill=white] (C2) circle (2pt);

\draw (V3) -- (V0);

\draw (V0) -- (V5)  node[midway] {$\times$};

\end{tikzpicture}
= + [q_{0345},1] \otimes [q_{0123},1]\,.
\end{equation}
Altogether the weighted symbol associated to this even hexagon has three terms:
\begin{equation}
T_{(0,1,2,3,4,5)}=
\begin{tikzpicture}[baseline=(current bounding box.center), scale=1.0, rotate=120]

\foreach \i in {0,...,5} {
    \coordinate (V\i) at ({cos(90 - \i*60 + 30)}, {sin(90 - \i*60 + 30)});
}

\foreach \i in {0,...,5} {
    \pgfmathtruncatemacro{\j}{mod(\i+1,6)}
    \draw (V\i) -- (V\j);
}

\foreach \i in {0,...,5} {
    \node at ({1.25*cos(90 - \i*60 + 30)}, {1.25*sin(90 - \i*60 + 30)}) {\(\i\)};
}

\draw (V0) -- (V5)  node[midway] {$\times$};

\end{tikzpicture}
= \begin{aligned}
&+ [q_{0125},1] \otimes [q_{2345},1]\\
&- [q_{0145},1] \otimes [1/q_{1234},1]\\
&+ [q_{0345},1] \otimes [q_{0123},1]\,.
\end{aligned}
\label{eq:evenhexagon}
\end{equation}

The odd hexagon $P=(1,2,3,4,5,6)$ has additional terms coming from the second term in~(\ref{eq:recursionodd}).  From the three quadrangulations we get
\begin{equation}
\begin{aligned}
\begin{tikzpicture}[baseline=(current bounding box.center), scale=1.0, rotate=180]

\foreach \i in {1,...,6} {
    \coordinate (V\i) at ({cos(90 - \i*60 + 30)}, {sin(90 - \i*60 + 30)});
}

\foreach \i in {1,...,6} {
    \pgfmathtruncatemacro{\j}{mod(\i,6)+1}
    \draw (V\i) -- (V\j);
}

\foreach \i in {1,...,6} {
    \node at ({1.25*cos(90 - \i*60 + 30)}, {1.25*sin(90 - \i*60 + 30)}) {\(\i\)};
}

\coordinate (C1) at (-0.375,0.216506);
\coordinate (C2) at (0.375,-0.216506);

\draw ($(C1)+(-3pt,-3pt)$) rectangle ($(C1)+(3pt,3pt)$);

\draw[postaction={decorate},decoration={markings,mark=at position 0.7 with {\arrow[scale=1.5]{Stealth}}},thick] (C1) -- (C2);

\fill (C2) circle (2pt);
\fill (C1) circle (2pt);

\draw (V1) -- (V4);

\draw (V6) -- (V1)  node[midway] {$\times$};

\end{tikzpicture}
&= + [1/q_{1456},1] \otimes [1/q_{1234},1]
+ [ 1/q_{1234}~  1/q_{1456}\color{black}, 2]\,,
\\
\begin{tikzpicture}[baseline=(current bounding box.center), scale=1.0, rotate=180]

\foreach \i in {1,...,6} {
    \coordinate (V\i) at ({cos(90 - \i*60 + 30)}, {sin(90 - \i*60 + 30)});
}

\foreach \i in {1,...,6} {
    \pgfmathtruncatemacro{\j}{mod(\i,6)+1}
    \draw (V\i) -- (V\j);
}

\foreach \i in {1,...,6} {
    \node at ({1.25*cos(90 - \i*60 + 30)}, {1.25*sin(90 - \i*60 + 30)}) {\(\i\)};
}

\coordinate (C1) at (0,0.433013);
\coordinate (C2) at (0,-0.433013);

\draw ($(C1)+(-3pt,-3pt)$) rectangle ($(C1)+(3pt,3pt)$);

\draw[postaction={decorate},decoration={markings,mark=at position 0.7 with {\arrow[scale=1.5]{Stealth}}},thick] (C1) -- (C2);

\fill (C1) circle (2pt);
\filldraw[fill=white] (C2) circle (2pt);

\draw (V2) -- (V5);

\draw (V6) -- (V1)  node[midway] {$\times$};

\end{tikzpicture}
&= - [1/q_{1256},1] \otimes [q_{2345},1]
- [ 1/q_{1256} ~ q_{2345}\color{black}, 2]\,,
\\
\begin{tikzpicture}[baseline=(current bounding box.center), scale=1.0, rotate=180]

\foreach \i in {1,...,6} {
    \coordinate (V\i) at ({cos(90 - \i*60 + 30)}, {sin(90 - \i*60 + 30)});
}

\foreach \i in {1,...,6} {
    \pgfmathtruncatemacro{\j}{mod(\i,6)+1}
    \draw (V\i) -- (V\j);
}

\foreach \i in {1,...,6} {
    \node at ({1.25*cos(90 - \i*60 + 30)}, {1.25*sin(90 - \i*60 + 30)}) {\(\i\)};
}

\coordinate (C1) at (0.375,0.216506);
\coordinate (C2) at (-0.375,-0.216506);

\draw ($(C1)+(-3pt,-3pt)$) rectangle ($(C1)+(3pt,3pt)$);

\draw[postaction={decorate},decoration={markings,mark=at position 0.7 with {\arrow[scale=1.5]{Stealth}}},thick] (C1) -- (C2);

\fill (C1) circle (2pt);
\fill (C2) circle (2pt);

\draw (V3) -- (V6);

\draw (V6) -- (V1)  node[midway] {$\times$};

\end{tikzpicture}
&= + [1/q_{1236},1] \otimes [1/q_{3456},1]
+ [ 1/q_{1236} ~ 1/q_{3456}\color{black}, 2]\,.
\end{aligned}
\end{equation}
Therefore, the total expression for this odd hexagon is
\begin{equation}
T_{(1,2,3,4,5,6)}=
\begin{tikzpicture}[baseline=(current bounding box.center), scale=1.0, rotate=180]

\foreach \i in {1,...,6} {
    \coordinate (V\i) at ({cos(90 - \i*60 + 30)}, {sin(90 - \i*60 + 30)});
}

\foreach \i in {1,...,6} {
    \pgfmathtruncatemacro{\j}{mod(\i,6)+1}
    \draw (V\i) -- (V\j);
}

\foreach \i in {1,...,6} {
    \node at ({1.25*cos(90 - \i*60 + 30)}, {1.25*sin(90 - \i*60 + 30)}) {\(\i\)};
}

\draw (V6) -- (V1)  node[midway] {$\times$};

\end{tikzpicture}
= \begin{aligned}
&+ [1/q_{1234} ~ 1/q_{1456},2]\\
&- [1/q_{1256} ~ q_{2345},2]\\
&+ [1/q_{1236} ~ 1/q_{3456},2]\\
&+ [1/q_{1236},1] \otimes [1/q_{3456},1] \\
&- [1/q_{1256},1] \otimes [q_{2345},1] \\
&+ [1/q_{1456},1] \otimes [1/q_{1234},1]\,.
\end{aligned}
\label{eq:oddhexagon}
\end{equation}

Finally we consider the even octagon $P = (0,1,\ldots,7)$.  Instead of writing down all 12 quadrangulations, it is sufficient to consider the 6 different choices of root quadrangle and recycle the results we already have for the hexagons.  There are 3 diagrams in which the root node is connected to a hexagon:
\begin{equation}
\begin{tikzpicture}[baseline=(current bounding box.center), scale=1.0, rotate=135]

\foreach \i in {0,...,7} {
    \coordinate (V\i) at ({cos(90 - \i*45 + 22.5)}, {sin(90 - \i*45 + 22.5)});
}

\foreach \i in {0,...,7} {
    \pgfmathtruncatemacro{\j}{mod(\i+1,8)}
    \draw (V\i) -- (V\j);
}

\foreach \i in {0,...,7} {
    \node at ({1.25*cos(90 - \i*45 + 22.5)}, {1.25*sin(90 - \i*45 + 22.5)}) {\(\i\)};
}

\coordinate (C1) at (-0.46194,0.46194);
\coordinate (C2) at (0.21776,-0.21776);
\%coordinate (C3) at (0.653281, 0);

\draw[rotate=45] ($(C1)+(-3pt,-3pt)$) rectangle ($(C1)+(3pt,3pt)$);

\draw[postaction={decorate},decoration={markings,mark=at position 0.7 with {\arrow[scale=1.5]{Stealth}}},thick] (C1) -- (C2);

\filldraw[fill=white] (C1) circle (2pt);
\fill (C2) circle (2pt);

\draw (V1) -- (V6);

\draw (V7) -- (V0)  node[midway] {$\times$};

\end{tikzpicture}
=
- [q_{0167},1] \otimes T_{(1,2,3,4,5,6)}\,,
\end{equation}

\begin{equation}
\begin{tikzpicture}[baseline=(current bounding box.center), scale=1.0, rotate=135]

\foreach \i in {0,...,7} {
    \coordinate (V\i) at ({cos(90 - \i*45 + 22.5)}, {sin(90 - \i*45 + 22.5)});
}

\foreach \i in {0,...,7} {
    \pgfmathtruncatemacro{\j}{mod(\i+1,8)}
    \draw (V\i) -- (V\j);
}

\foreach \i in {0,...,7} {
    \node at ({1.25*cos(90 - \i*45 + 22.5)}, {1.25*sin(90 - \i*45 + 22.5)}) {\(\i\)};
}

\coordinate (C1) at (0,0.653281);
\coordinate (C2) at (0,-0.30796);

\draw[rotate=-135] ($(C1)+(-3pt,-3pt)$) rectangle ($(C1)+(3pt,3pt)$);

\draw[postaction={decorate},decoration={markings,mark=at position 0.7 with {\arrow[scale=1.5]{Stealth}}},thick] (C1) -- (C2);

\filldraw[fill=white] (C1) circle (2pt);
\filldraw[fill=white] (C2) circle (2pt);

\draw (V2) -- (V7);

\draw (V7) -- (V0)  node[midway] {$\times$};

\end{tikzpicture}
= - [q_{0127},1] \otimes T_{(2,3,4,5,6,7)}\,,
\end{equation}

\begin{equation}
\begin{tikzpicture}[baseline=(current bounding box.center), scale=1.0, rotate=135]

\foreach \i in {0,...,7} {
    \coordinate (V\i) at ({cos(90 - \i*45 + 22.5)}, {sin(90 - \i*45 + 22.5)});
}

\foreach \i in {0,...,7} {
    \pgfmathtruncatemacro{\j}{mod(\i+1,8)}
    \draw (V\i) -- (V\j);
}

\foreach \i in {0,...,7} {
    \node at ({1.25*cos(90 - \i*45 + 22.5)}, {1.25*sin(90 - \i*45 + 22.5)}) {\(\i\)};
}

\coordinate (C2) at (0.30796,0);
\coordinate (C3) at (-0.653281,0);

\draw[rotate=-135] ($(C3)+(-3pt,-3pt)$) rectangle ($(C3)+(3pt,3pt)$);

\draw[postaction={decorate},decoration={markings,mark=at position 0.7 with {\arrow[scale=1.5]{Stealth}}},thick] (C3) -- (C2);

\filldraw[fill=white] (C2) circle (2pt);
\filldraw[fill=white] (C3) circle (2pt);

\draw (V5) -- (V0);

\draw (V7) -- (V0)  node[midway] {$\times$};

\end{tikzpicture}
\begin{aligned}
= -[q_{0567},1] \otimes T_{(0,1,2,3,4,5)}\,,
\end{aligned}
\end{equation}

and another 3 in which the root node is connected to two quadrilaterals:

\begin{equation}
\begin{tikzpicture}[baseline=(current bounding box.center), scale=1.0, rotate=135]

\foreach \i in {0,...,7} {
    \coordinate (V\i) at ({cos(90 - \i*45 + 22.5)}, {sin(90 - \i*45 + 22.5)});
}

\foreach \i in {0,...,7} {
    \pgfmathtruncatemacro{\j}{mod(\i+1,8)}
    \draw (V\i) -- (V\j);
}

\foreach \i in {0,...,7} {
    \node at ({1.25*cos(90 - \i*45 + 22.5)}, {1.25*sin(90 - \i*45 + 22.5)}) {\(\i\)};
}

\coordinate (C1) at (0.46194,0.46194);
\coordinate (C2) at (-0.326641,0.135299);
\coordinate (C3) at (0,-0.653281);

\draw[rotate=-135] ($(C2)+(-3pt,-3pt)$) rectangle ($(C2)+(3pt,3pt)$);

\draw[postaction={decorate},decoration={markings,mark=at position 0.7 with {\arrow[scale=1.5]{Stealth}}},thick] (C2) -- (C1);
\draw[postaction={decorate},decoration={markings,mark=at position 0.7 with {\arrow[scale=1.5]{Stealth}}},thick] (C2) -- (C3);

\filldraw[fill=white] (C1) circle (2pt);
\filldraw[fill=white] (C2) circle (2pt);
\fill (C3) circle (2pt);

\draw (V3) -- (V0);
\draw (V3) -- (V6);

\draw (V7) -- (V0)  node[midway] {$\times$};

\end{tikzpicture}
=(-1)^2[q_{0367},1] \otimes ([q_{0123},1] \star [1/q_{3456},1])\,,
\end{equation}

\begin{equation}
\begin{tikzpicture}[baseline=(current bounding box.center), scale=1.0, rotate=135]

\foreach \i in {0,...,7} {
    \coordinate (V\i) at ({cos(90 - \i*45 + 22.5)}, {sin(90 - \i*45 + 22.5)});
}

\foreach \i in {0,...,7} {
    \pgfmathtruncatemacro{\j}{mod(\i+1,8)}
    \draw (V\i) -- (V\j);
}

\foreach \i in {0,...,7} {
    \node at ({1.25*cos(90 - \i*45 + 22.5)}, {1.25*sin(90 - \i*45 + 22.5)}) {\(\i\)};
}

\coordinate (C1) at (0.653281,0);
\coordinate (C2) at (-0.135299,0.326641);
\coordinate (C3) at (-0.46194,-0.46194);

\draw[rotate=-135] ($(C2)+(-3pt,-3pt)$) rectangle ($(C2)+(3pt,3pt)$);

\draw[postaction={decorate},decoration={markings,mark=at position 0.7 with {\arrow[scale=1.5]{Stealth}}},thick] (C2) -- (C1);
\draw[postaction={decorate},decoration={markings,mark=at position 0.7 with {\arrow[scale=1.5]{Stealth}}},thick] (C2) -- (C3);

\fill (C1) circle (2pt);
\filldraw[fill=white] (C2) circle (2pt);
\filldraw[fill=white] (C3) circle (2pt);

\draw (V4) -- (V1);
\draw (V4) -- (V7);

\draw (V7) -- (V0)  node[midway] {$\times$};

\end{tikzpicture}
= (-1)^2[q_{0147},1] \otimes ([1/q_{1234},1] \star [q_{4567}, 1])\,,\\
\end{equation}

\begin{equation}
\begin{tikzpicture}[baseline=(current bounding box.center), scale=1.0, rotate=135]

\foreach \i in {0,...,7} {
    \coordinate (V\i) at ({cos(90 - \i*45 + 22.5)}, {sin(90 - \i*45 + 22.5)});
}

\foreach \i in {0,...,7} {
    \pgfmathtruncatemacro{\j}{mod(\i+1,8)}
    \draw (V\i) -- (V\j);
}

\foreach \i in {0,...,7} {
    \node at ({1.25*cos(90 - \i*45 + 22.5)}, {1.25*sin(90 - \i*45 + 22.5)}) {\(\i\)};
}

\coordinate (C1) at (0.46194,0.46194);
\coordinate (C2) at (0,0);
\coordinate (C3) at (-0.46194,-0.46194);

\draw[rotate=-135] ($(C2)+(-3pt,-3pt)$) rectangle ($(C2)+(3pt,3pt)$);

\draw[postaction={decorate},decoration={markings,mark=at position 0.85 with {\arrow[scale=1.5]{Stealth}}},thick] (C2) -- (C1);
\draw[postaction={decorate},decoration={markings,mark=at position 0.85 with {\arrow[scale=1.5]{Stealth}}},thick] (C2) -- (C3);

\filldraw[fill=white] (C1) circle (2pt);
\filldraw[fill=white] (C2) circle (2pt);
\filldraw[fill=white] (C3) circle (2pt);

\draw (V0) -- (V3);
\draw (V4) -- (V7);

\draw (V7) -- (V0)  node[midway] {$\times$};

\end{tikzpicture}
= (-1)^3 [q_{0347},1] \otimes ([q_{0123},1] \star [q_{4567},1])\,,
\end{equation}
where we have pulled out the overall sign of each term (which gets a contribution of $-1$ from each white vertex) and remind the reader that~(\ref{eq:stardef}) gives
\begin{align}
    [\varphi_1, 1] \star [\varphi_2,1] = [\varphi_1\,\varphi_2,2] + [\varphi_1,1] \otimes [\varphi_2,1] + [\varphi_2,1] \otimes [\varphi_1,1]\,.
\end{align}
Altogether there are a total of $6+3+3+3+3+3=21$ terms.  Odd octagons are considerably more complicated due to the second term in~(\ref{eq:recursionodd}), and have a total of 48 terms.

\subsection{Quadrangular Polylogarithms}
\label{sec:quadpoly}

Rudenko defines quadrangular polylogarithms for any alternating $(2n{+}2)$-gon $P$ and any integer $k \ge 0$ such that
\begin{align}
\label{eq:QLidef}
    \QLi^{\rm R}_{k}(P) =\Li^{\rm R}_k(T_P)\,,
\end{align}
where $\Li^{\rm R}_k$ distributes across words according to
\begin{align}
\label{eq:distribute}
    \Li^{\rm R}_k(W_1 \pm W_2 \pm \cdots) = \Li^{\rm R}_k(W_1) \pm \Li^{\rm R}_k(W_2) \pm \cdots\,.
\end{align}
We will not provide here the general definition of $\Li^{\rm R}_k(W)$ since our formula for dS wavefunction coefficients only requires the cases $k=0,1$ which we describe momentarily.

At this point we break with Rudenko's conventions and notation in three ways.  First of all, it is clear that Rudenko's construction leads to functionally distinct expressions for $\QLi^{\rm R}_k(T_P)$ depending only on whether $P$ is even or odd. We will find it convenient to make the distinction explicit by defining two different functions $\QLi^\pm$. Secondly, if $P$ is a $(2n{+}2)$-gon, then Rudenko's $\Li^{\rm R}_k(T_P)$ has weight $n+k$, not $k$, but we choose to record the weight of $\QLi$ in its subscript, as is conventional for the classical polylogarithm function $\Li_k$.  Third, as emphasized above, Rudenko's definition treats $\QLi$ not as a specific function but as an equivalence class of functions modulo products of functions of lower weight.  Instead, we shall interpret~(\ref{eq:QLidef}) as an exact equality of functions with the right-hand side given for $k=0$ by~(\ref{eq:distribute}) with
\begin{align}
    \Li^{\rm R}_0([\varphi_1, m_1] \otimes \cdots \otimes [\varphi_d,m_d]) = \Li_{m_1,\ldots,m_d}(\varphi_1,\ldots,\varphi_d)
\end{align}
in terms of the standard multiple polylogarithm function~\cite{goncharov1995polylogarithms}, defined for positive integers $n_i$ and complex numbers $|z_i|<1$ by the power series
\begin{align}\label{eq: MPL_series_expansion}
    \Li_{m_1,m_2,\dots, m_d}(z_1,z_2,\dots,z_d)=\sum_{0<n_1<n_2<\dots<n_d}\frac{z_1^{n_1} z_2^{n_2}\dots z_d^{n_d}}{n_1^{m_1}n_2^{m_2}\dots n_d^{m_d}}\,.
\end{align}
Here we use the convention of~\cite{Goncharov:1998kja,Duhr:2019tlz} which differs from that of~\cite{rudenko2023goncharov} by the reversal of arguments. For $k=1$ we have
\begin{align}
\label{eq:keq1}
    \Li_1^{\rm R}([\varphi_1, m_1] \otimes \cdots \otimes [\varphi_d,m_d]) = - \sum_{i=1}^d
    m_i\,\Li_{m_1,\ldots,m_i+1,\ldots,m_d}(\varphi_1,\ldots,\varphi_d)\,.
\end{align}
Altogether, we define, for $k \ge n-1$, the \emph{even and odd quadrangular polylogarithm functions} $\QLi_k^{\pm}$ by
\begin{align}
    \QLi_k^+(i_1,i_2,\ldots,i_{2n}) &= \QLi_{k-n+1}^{\rm R}(0,1,\ldots,2n{-}1)\rvert_{z_j \to z_{i_{j+1}}}\,, \label{eq:QLi1plus} \\
    \QLi_k^-(i_1,i_2,\ldots,i_{2n}) &= \QLi_{k-n+1}^{\rm R}(1,2,\ldots,2n)\rvert_{z_j \to z_{i_j}}\,.
\end{align}

\subsubsection{Examples}

For $n = 2$ and $k \ge 1$ we have
\begin{align}\label{eq:QLiquad}
    \QLi^{+}_k(a,b,c,d) &= (-1)^k \Li_k(q_{abcd})\,,\\
    \QLi^{-}_k(a,b,c,d) &= (-1)^{k+1} \Li_k(1/q_{abcd})\,.
\end{align}
For $n = 3$ and $k=2$ we have
\begin{align}
    \QLi^{+}_2(a,b,c,d,e,f) &= \Li_{1,1}(q_{abcf}, q_{cdef}) - \Li_{1,1}(q_{abef},1/q_{bcde})\nonumber \\
    &\qquad\qquad+ \Li_{1,1}(q_{adef},q_{abcd})\,, \\
    \QLi^{-}_2(a,b,c,d,e,f) &= \Li_2(1/( q_{abcd}\, q_{adef})) + \Li_{1,1}(1/q_{abcf}, 1/q_{cdef})\nonumber \\
    &\qquad\qquad- \Li_{1,1}(1/q_{abef},q_{bcde}) + \Li_{1,1}(1/q_{adef}, 1/q_{abcd})
    \label{eq:example4}
\end{align}
while for $k=3$ it follows from~(\ref{eq:keq1}) that
\begin{equation}
\begin{aligned}
    \QLi_3^+(a,b,c,d,e,f) =
    &-\Li_{2,1}(q_{abcf}, q_{cdef}) -  \Li_{1,2}(q_{abcf}, q_{cdef})\\
    &+ \Li_{2,1}(q_{abef},1/q_{bcde}) + \Li_{1,2}(q_{abef},1/q_{bcde})\\
    &- \Li_{2,1}(q_{adef},q_{abcd}) - \Li_{1,2}(q_{adef},q_{abcd})\,,\\
    \QLi_3^-(a,b,c,d,e,f) =
    &-2 \Li_3(1/(q_{abcd}\, q_{adef})) \\
    &- \Li_{2,1}(1/q_{abcf}, 1/q_{cdef}) - \Li_{1,2}(1/q_{abcf}, 1/q_{cdef}) \\
    &+ \Li_{2,1}(1/q_{abef},q_{bcde}) + \Li_{1,2}(1/q_{abef},q_{bcde}) \\
    &-  \Li_{2,1}(1/q_{adef}, 1/q_{abcd}) -  \Li_{1,2}(1/q_{adef}, 1/q_{abcd})\,.
\end{aligned}
\end{equation}
Note that although the expression~(\ref{eq:oddhexagon}) has six terms, the first three are all equal to each other (up to sign) when evaluated on~(\ref{eq:qdef}) and two of them cancel leaving only the first term shown in~(\ref{eq:example4}).

For $n = 4$ and $k=3$ we have
\begin{equation}
\begin{aligned}
\QLi_3^+(a,b,c, &d,e,f,g,h)\\
&= \Li_{1,2}(q_{abeh},q_{efgh}/q_{bcde})
-\Li_{1,2}(q_{abgh},1/(q_{bcde}\, q_{befg}))\\
&-\Li_{1,2}(q_{adeh},q_{abcd}\,q_{efgh})
+\Li_{1,2}(q_{adgh},q_{abcd}/q_{defg})\\
&-\Li_{1,1,1}(q_{abch},q_{cdeh},q_{efgh})
+\Li_{1,1,1}(q_{abch},q_{cdgh},1/q_{defg})\\
&-\Li_{1,1,1}(q_{abch},q_{cfgh},q_{cdef})
+\Li_{1,1,1}(q_{abeh},1/q_{bcde},q_{efgh})\\
&+\Li_{1,1,1}(q_{abeh},q_{efgh},1/q_{bcde})
-\Li_{1,1,1}(q_{abgh},1/q_{bcdg},1/q_{defg})\\
&+\Li_{1,1,1}(q_{abgh},1/q_{bcfg},q_{cdef})
-\Li_{1,1,1}(q_{abgh},1/q_{befg},1/q_{bcde})\\
&-\Li_{1,1,1}(q_{adeh},q_{abcd},q_{efgh})
-\Li_{1,1,1}(q_{adeh},q_{efgh},q_{abcd})\\
&+\Li_{1,1,1}(q_{adgh},q_{abcd},1/q_{defg})
+\Li_{1,1,1}(q_{adgh},1/q_{defg},q_{abcd})\\
&-\Li_{1,1,1}(q_{afgh},q_{abcf},q_{cdef})
+\Li_{1,1,1}(q_{afgh},q_{abef},1/q_{bcde})\\
&-\Li_{1,1,1}(q_{afgh},q_{adef},q_{abcd})\,.
\end{aligned}
\end{equation}
The explicit expressions for $\QLi_3^-$, $\QLi_4^+$ and $\QLi_4^-$ at $n=4$ are easy to write down using the rules we have given above and have 26, 53, and 66 terms respectively.

\subsection{Coproduct}

Multiple polylogarithm functions are elements of a Hopf algebra~\cite{goncharov1995polylogarithms}, with a coproduct that we denote $\Delta$. We shall only make use of one of its components: when acting on a multiple polylogarithm function $F_n$ of weight $n$, the coproduct component $\Delta_{n-1,1}$ encodes the differential $dF_n$.  Specifically, the formula
\begin{align}
    dF_n = F_{n-1}^{(1)} d \log f_1 + F_{n-1}^{(2)} d \log f_2 + \cdots\,,
\end{align}
which expresses the total differential of $F_n$ as a sum of $d\log$ forms having coefficients $F_{n-1}^{(i)}$ that are multiple polylogarithms of weight $n{-}1$, should be understood as synonymous with the coproduct formula
\begin{align}
    \Delta_{n-1,1} F_n = F_{n-1}^{(1)} \otimes \log f_1 + F_{n-1}^{(2)} \otimes \log f_2 + \cdots\,.
\end{align}
In particular, the main theme of our proof in Section~\ref{sec:proof} relies on the fact that if two functions $F_n$ and $G_n$ have the same $\Delta_{n-1,1}$ coproduct component, then they differ from each other only by a constant.

Rudenko provides, in Theorem~5.5 of~\cite{rudenko2023goncharov}, a formula for the coproduct of quadrangular polylogarithms modulo products obtained by using their recursive definition in terms of quadrangulations. It is straightforward to generalize that argument to obtain a formula for the full coproduct of our $\QLi^\pm$ functions.  The results are rather lengthy so we display here only the two specific formulas that we shall need in the following sections: 
\begin{equation}\label{eq:DeltaQLi}
    \begin{aligned}
        &\Delta_{n-1,1} \qli^{\pm}_{n}(0,\ldots,2n{-}1) =
        \mp \qli^{\pm}_{n-1}(0,\ldots,2n{-}1) \otimes \log\left( \prod_{i=1}^{n-1} q_{0,2i-1,2i,2i+1} \right) \\
        &\qquad\qquad\qquad\qquad+ \sum_{\substack{0 < i < j < 2n-1 \\ i \text{ odd } j \text{ even }}} \sum_{ \substack{m_\bullet \in \{0,1\} \\ m_1+m_2+m_3 = 1} } T_{0,i,j,2n-1}^{\pm,m_1,m_2,m_3} \otimes \qli^{\pm}_1(0,i,j,2n{-}1),
    \end{aligned}
\end{equation}
and
\begin{equation}\label{eq:DeltaQLi0}
    \begin{aligned}
        \Delta_{n-2,1} \qli^{\pm}_{n-1}(0,\ldots,2n{-}1)
        &= \sum_{\substack{1 < i < j < 2n \\ i \text{ odd } j \text{ even }}} T_{0,i,j,2n-1}^{\pm,0,0,0} \otimes \qli^{\pm}_1(0,i,j,2n{-}1) \,, \\
    \end{aligned}
\end{equation}
where
\begin{multline}\label{eq:defT}
    T_{0,i,j,2n-1}^{\pm,m_1,m_2,m_3} = \QLi^{\pm}_{(i-1)/2+m_1}(0,\ldots,i)\\
    \times \QLi^{\mp}_{(j-i-1)/2+m_2}(i,\ldots,j) \QLi^{\pm}_{(2n-j)/2+m_3-1}(j,\ldots,2n{-}1)\,,
\end{multline}
and it is understood that
\begin{align}\label{eq:qli2}
    \QLi^{\pm}_0(i,j) &= 1\,,\\
    \QLi^{\pm}_1(i,j) &= 0\,.
\end{align}

\section{Physics: Recursion Relation for Chain Wavefunctions}
\label{sec:recursion}

A symbol-level recursion for dS wavefunction coefficients was written down several years ago in~\cite{Hillman:2019wgh}, but our starting point is instead the recursion given in~\cite{He:2024olr}. The two recursions have a different flavor: the former is phrased in terms of discontinuities and builds the symbol from right to left while the latter is phrased in terms of derivatives and builds the symbol from left to right.  The proof of our main result in Section~\ref{sec:proof} relies on the remarkable fact that after a little rewriting, the latter describes a structure almost identical to that encoded in Rudenko's coproduct formula~\eqref{eq:DeltaQLi}.

In this section we show that the wavefunction coefficient for the $n$-site chain
\begin{align}
    \psi_{n}(X_1,X_2,\ldots,X_n; Y_{1,2},Y_{2,3},\ldots,Y_{n-1,n}) =
\begin{tikzpicture}[baseline=(current bounding box.center),scale=1.3]
		\draw[line width=1.5pt] (0,0)--(1.75,0);
		\draw[line width=1.5pt] (2.25,0)--(4,0);
		\draw[fill=black] (0,0) circle [radius=0.075];
		\draw[fill=black] (1.5-0.2,0) circle [radius=0.075];
		\draw[fill=black] (2.5+0.2,0) circle [radius=0.075];
		\draw[fill=black] (4,0) circle [radius=0.075];
		\draw[fill=black] (1.9,0) circle [radius=0.02];
		\draw[fill=black] (2,0) circle [radius=0.02];
		\draw[fill=black] (2.1,0) circle [radius=0.02];
		\node[below] at (0,-0.1) {{$X_1$}};
		\node[below] at (4,-0.1) {{$X_n$}};
		\node[below] at (1.5-0.25,-0.1) {{$X_{2}$}};
		\node[below] at (2.5+0.25,-0.1) {{$X_{n-1}$}};
		\node[above] at (0.625,0.01) {{${Y}_{1,2}$}};
        \node[above] at (3.375,0.01) {{$Y_{n-1,n}$}};
\end{tikzpicture}
\label{eq:psin}
\end{align}
satisfies the simple coproduct recursion~(\ref{eq:simplifiedDelta}) when appropriately expressed in terms of cross-ratios of $2n{+}1$ points $z_i \in \mathbb{P}^1$. Our starting point is equation (5.15) of~\cite{He:2024olr}, which implies that for $n \geq 2$, the symbol of $\psi_{n}$ can be expressed recursively in terms of $(n{-}1)$-site chain wavefunction coefficients with shifted arguments:
\begin{equation}\label{eq:dchain}
\begin{aligned}
&\Delta_{n-1,1}\psi_{n}
=\left(\raisebox{-1.5em}{\begin{tikzpicture}[scale=1.3]
		\draw[line width=1.5pt] (0,0)--(1,0);
		\draw[line width=1.5pt] (1,0)--(1.5,0);
		\draw[line width=1.5pt] (2,0)--(2.5,0);
		\draw[fill=black] (0,0) circle [radius=0.075];
		\draw[fill=black] (1,0) circle [radius=0.075];
		\draw[fill=black] (2.5,0) circle [radius=0.075];
		\node[] at (0,-0.3) {\small{$X_2+Y_{1,2}$}};
		\node[] at (1,-0.3) {\small{$X_3$}};
		\node[] at (2.5,-0.3) {\small{$X_n$}};
		\draw[fill=black] (1.5+0.5/4,0) circle [radius=0.025];
		\draw[fill=black] (1.5+2*0.5/4,0) circle [radius=0.025];
		\draw[fill=black] (1.5+3*0.5/4,0) circle [radius=0.025];
	\end{tikzpicture}}{-}\raisebox{-1.5em}{\begin{tikzpicture}[scale=1.3]
		\draw[line width=1.5pt] (0,0)--(1,0);
		\draw[line width=1.5pt] (1,0)--(1.5,0);
		\draw[line width=1.5pt] (2,0)--(2.5,0);
		\draw[fill=black] (0,0) circle [radius=0.075];
		\draw[fill=black] (1,0) circle [radius=0.075];
		\draw[fill=black] (2.5,0) circle [radius=0.075];
		\node[] at (0,-0.3) {\small{$X_{1}{+}X_{2}$}};
		\node[] at (1,-0.3) {\small{$X_3$}};
		\node[] at (2.5,-0.3) {\small{$X_n$}};
		\draw[fill=black] (1.5+0.5/4,0) circle [radius=0.025];
		\draw[fill=black] (1.5+2*0.5/4,0) circle [radius=0.025];
		\draw[fill=black] (1.5+3*0.5/4,0) circle [radius=0.025];
	\end{tikzpicture}}\right)\otimes \log{\frac{X_1^-}{X_1^+}}\\
&+\sum_{i=2}^{n-1}\left({\begin{tikzpicture}[baseline=(current bounding box.center),scale=1.3]
		\draw[line width=1.5pt] (0,0)--(0.5,0);
		\draw[line width=1.5pt] (1,0)--(3,0);
		\draw[line width=1.5pt] (3.5,0)--(4,0);
		\draw[fill=black] (0,0) circle [radius=0.075];
		\draw[fill=black] (1.5-0.2,0) circle [radius=0.075];
		\draw[fill=black] (2.5+0.2,0) circle [radius=0.075];
		\draw[fill=black] (4,0) circle [radius=0.075];
		\draw[fill=black] (0.5+0.5/4,0) circle [radius=0.02];
		\draw[fill=black] (0.5+2*0.5/4,0) circle [radius=0.02];
		\draw[fill=black] (0.5+3*0.5/4,0) circle [radius=0.02];
		\draw[fill=black] (3+0.5/4,0) circle [radius=0.02];
		\draw[fill=black] (3+2*0.5/4,0) circle [radius=0.02];
		\draw[fill=black] (3+3*0.5/4,0) circle [radius=0.02];
		\node[] at (0,-0.3) {\small{$X_1$}};
		\node[] at (4,-0.3) {\small{$X_n$}};
		\node[] at (1.5-0.25,-0.3) {\small{$\tilde{X}_{i-1}$}};
		\node[] at (2.5+0.25,-0.3) {\small{$\tilde{X}_{i+1}$}};
		\node[above] at (2,0.01) {\small{$\tilde{Y}_{i{-}1,i{+}1}$}};
	\end{tikzpicture}} \otimes \log{\frac{X_i^{+-}X_i^{-+}}{X_i^{++}X_i^{--}}} \right. \\
&\left.-{\begin{tikzpicture}[baseline=(current bounding box.center),scale=1.3]
		\draw[line width=1.5pt] (0,0)--(0.5,0);
		\draw[line width=1.5pt] (1,0)--(2,0);
		\draw[line width=1.5pt] (2.5,0)--(3,0);
		\draw[fill=black] (0,0) circle [radius=0.075];
		\draw[fill=black] (1.5,0) circle [radius=0.075];
		\draw[fill=black] (3,0) circle [radius=0.075];
		\draw[fill=black] (0.5+0.5/4,0) circle [radius=0.025];
		\draw[fill=black] (0.5+2*0.5/4,0) circle [radius=0.025];
		\draw[fill=black] (0.5+3*0.5/4,0) circle [radius=0.025];
		\draw[fill=black] (2+0.5/4,0) circle [radius=0.025];
		\draw[fill=black] (2+2*0.5/4,0) circle [radius=0.025];
		\draw[fill=black] (2+3*0.5/4,0) circle [radius=0.025];
		\node[] at (0,-0.3) {\small{$X_1$}};
		\node[] at (3,-0.3) {\small{$X_n$}};
		\node[] at (1.5,-0.3) {\small{$X_{i}{+}X_{i{+}1}$}};
		\node[above] at (1.1,0) {\small{$Y_{i{-}1,i}$}};
		\node[above] at (2.1,0) {\small{$Y_{i{+}1,i{+}2}$}};
	\end{tikzpicture}}\otimes \log{\frac{X_i^{+-}}{X_i^{++}}}-{\begin{tikzpicture}[baseline=(current bounding box.center),scale=1.3]
		\draw[line width=1.5pt] (0,0)--(0.5,0);
		\draw[line width=1.5pt] (1,0)--(2,0);
		\draw[line width=1.5pt] (2.5,0)--(3,0);
		\draw[fill=black] (0,0) circle [radius=0.075];
		\draw[fill=black] (1.5,0) circle [radius=0.075];
		\draw[fill=black] (3,0) circle [radius=0.075];
		\draw[fill=black] (0.5+0.5/4,0) circle [radius=0.025];
		\draw[fill=black] (0.5+2*0.5/4,0) circle [radius=0.025];
		\draw[fill=black] (0.5+3*0.5/4,0) circle [radius=0.025];
		\draw[fill=black] (2+0.5/4,0) circle [radius=0.025];
		\draw[fill=black] (2+2*0.5/4,0) circle [radius=0.025];
		\draw[fill=black] (2+3*0.5/4,0) circle [radius=0.025];
		\node[] at (0,-0.3) {\small{$X_1$}};
		\node[] at (3,-0.3) {\small{$X_n$}};
		\node[] at (1.5,-0.3) {\small{$X_{i{-}1}{+}X_i$}};
		\node[above] at (1.1,0) {\small{$Y_{i{-}2,i{-}1}$}};
		\node[above] at (2.1,0) {\small{$Y_{i,i{+}1}$}};
	\end{tikzpicture}}\otimes \log{\frac{X_i^{-+}}{X_i^{++}}}
    \vphantom{\begin{tikzpicture}[baseline=(current bounding box.center),scale=1.3]
		\draw[line width=1.5pt] (0,0)--(0.5,0);
		\draw[line width=1.5pt] (1,0)--(3,0);
		\draw[line width=1.5pt] (3.5,0)--(4,0);
		\draw[fill=black] (0,0) circle [radius=0.075];
		\draw[fill=black] (1.5-0.2,0) circle [radius=0.075];
		\draw[fill=black] (2.5+0.2,0) circle [radius=0.075];
		\draw[fill=black] (4,0) circle [radius=0.075];
		\draw[fill=black] (0.5+0.5/4,0) circle [radius=0.02];
		\draw[fill=black] (0.5+2*0.5/4,0) circle [radius=0.02];
		\draw[fill=black] (0.5+3*0.5/4,0) circle [radius=0.02];
		\draw[fill=black] (3+0.5/4,0) circle [radius=0.02];
		\draw[fill=black] (3+2*0.5/4,0) circle [radius=0.02];
		\draw[fill=black] (3+3*0.5/4,0) circle [radius=0.02];
		\node[] at (0,-0.3) {\small{$X_1$}};
		\node[] at (4,-0.3) {\small{$X_n$}};
		\node[] at (1.5-0.25,-0.3) {\small{$\tilde{X}_{i-1}$}};
		\node[] at (2.5+0.25,-0.3) {\small{$\tilde{X}_{i+1}$}};
		\node[above] at (2,0.01) {\small{$\tilde{Y}_{i{-}1,i{+}1}$}};
	\end{tikzpicture}}
    \right)\\
& + \left(\raisebox{-1.5em}{\begin{tikzpicture}[scale=1.3]
		\draw[line width=1.5pt] (0.5,0)--(-1,0);
		\draw[line width=1.5pt] (-1,0)--(-1.5,0);
		\draw[line width=1.5pt] (-2,0)--(-2.5,0);
		\draw[fill=black] (0.5,0) circle [radius=0.075];
		\draw[fill=black] (-1,0) circle [radius=0.075];
		\draw[fill=black] (-2.5,0) circle [radius=0.075];
		\node[] at (0.5,-0.3) {\small{$X_{n{-}1}{+}Y_{n{-}1,n}$}};
		\node[] at (-1,-0.3) {\small{$X_{n-2}$}};
		\node[] at (-2.5,-0.3) {\small{$X_1$}};
		\draw[fill=black] (-1.5-0.5/4,0) circle [radius=0.025];
		\draw[fill=black] (-1.5-2*0.5/4,0) circle [radius=0.025];
		\draw[fill=black] (-1.5-3*0.5/4,0) circle [radius=0.025];
	\end{tikzpicture}}
-\raisebox{-1.5em}{\begin{tikzpicture}[scale=1.3]
		\draw[line width=1.5pt] (0,0)--(-1,0);
		\draw[line width=1.5pt] (-1,0)--(-1.5,0);
		\draw[line width=1.5pt] (-2,0)--(-2.5,0);
		\draw[fill=black] (0,0) circle [radius=0.075];
		\draw[fill=black] (-1,0) circle [radius=0.075];
		\draw[fill=black] (-2.5,0) circle [radius=0.075];
		\node[] at (0,-0.3) {\small{$X_{n-1}{+}X_n$}};
		\node[] at (-1,-0.3) {\small{$X_{n-2}$}};
		\node[] at (-2.5,-0.3) {\small{$X_1$}};
		\draw[fill=black] (-1.5-0.5/4,0) circle [radius=0.025];
		\draw[fill=black] (-1.5-2*0.5/4,0) circle [radius=0.025];
		\draw[fill=black] (-1.5-3*0.5/4,0) circle [radius=0.025];
	\end{tikzpicture}}\right)\otimes \log{\frac{X_n^-}{X_n^+}}\,.
\end{aligned}
\end{equation}
The shifted arguments in the first term in the sum are
\begin{equation}\label{eq:tildeXY}
\begin{aligned}
\tilde{X}_{i-1} &= X_{i-1} + \tfrac{1}{2} X^{+-}_i\,,\\
\tilde{X}_{i+1} &= X_{i+1} + \tfrac{1}{2} X^{-+}_i\,,\\
\tilde{Y}_{i-1,i+1} &= - \tfrac{1}{2} X^{--}_i\,,
\end{aligned}
\end{equation}
and we used the notation
\begin{align}
\label{eq:xpmxpmdef}
X_i^{\pm \pm} = X_i \pm Y_{i-1,i} \pm Y_{i,i+1}\,,
\end{align}
with the understanding that $Y_{0,1} = Y_{n,n+1} = 0$.

Following~\cite{Paranjape:2026htn} and the examples discussed in Section~\ref{sec:prelude}, we now embed the variables into $\Gr(2,2n{+}1)$ via
\begin{equation}\label{eq:Cpn}
\begin{aligned}
    C_{n}(X;Y) &=\begin{pmatrix}
1&1&1&1&1&\cdots &1 &1 &1&0\\
        z_1&z_2&z_3&z_4&z_5&\cdots &z_{2n-2}&z_{2n-1} &z_{2n}&1
    \end{pmatrix},
\end{aligned}
\end{equation}
where $z_1 = 0$,
\begin{align}
    z_{2i} = \sum_{k=1}^i X_k + Y_{i,i+1}\,, \qquad
    z_{2i+1} = \sum_{k=1}^i X_k - Y_{i,i+1}\,, \quad
    1 \le i < n-1\,,
\end{align}
and $z_{2n} = \sum_{i=1}^n X_i$. Henceforth we view the wavefunction coefficient $\psi_n$ as a function of the $z_i$, instead of as a function of the $X$'s and $Y$'s as in~(\ref{eq:psin}).

Our next task is to rewrite the equation~(\ref{eq:dchain}) in this notation. The arguments of the logarithms in~(\ref{eq:dchain}) can then be expressed in terms of the cross-ratio defined in~(\ref{eq:qdef}) as
\begin{equation}
\label{eq:XXtoq}
\begin{aligned}
    \frac{X_i^{+-}}{X_i^{++}}&=q_{2i-1,2i{+}1,2n+1,2i}\,,&&i=1,2\ldots,n{-}1\,,\\
    \frac{X_i^{-+}}{X_i^{++}}&=q_{2i-2,2i,2i-1,2n+1}\,, &&i=2,3\ldots,n\,,\\
    \frac{X_i^{+-}X_i^{-+}}{X_i^{++}X_i^{--}}&=q_{2i-1,2i{+}1,2i-2,2i}\,,&&i=2,3,\ldots,n{-}1\,.
\end{aligned}
\end{equation}
Now comes the most remarkable step: when expressed in these variables, the shifted arguments appearing in each one of the $(n{-}1)$-site chain wavefunction coefficients appearing on the right-hand side of~(\ref{eq:dchain}) are precisely those associated to the matrix~(\ref{eq:Cpn}) but with two adjacent columns removed.  Let us demonstrate how this works for the first term in the sum,
\begin{equation}
\psi_{n-1}(\ldots,\tilde{X}_{i-1},\tilde{X}_{i+1},\ldots;\ldots,\tilde{Y}_{i-1,i+1},\ldots)=
\begin{tikzpicture}[baseline=(current bounding box.center),scale=1.3]
		\draw[line width=1.5pt] (0,0)--(0.5,0);
		\draw[line width=1.5pt] (1,0)--(3,0);
		\draw[line width=1.5pt] (3.5,0)--(4,0);
		\draw[fill=black] (0,0) circle [radius=0.075];
		\draw[fill=black] (1.5-0.2,0) circle [radius=0.075];
		\draw[fill=black] (2.5+0.2,0) circle [radius=0.075];
		\draw[fill=black] (4,0) circle [radius=0.075];
		\draw[fill=black] (0.5+0.5/4,0) circle [radius=0.02];
		\draw[fill=black] (0.5+2*0.5/4,0) circle [radius=0.02];
		\draw[fill=black] (0.5+3*0.5/4,0) circle [radius=0.02];
		\draw[fill=black] (3+0.5/4,0) circle [radius=0.02];
		\draw[fill=black] (3+2*0.5/4,0) circle [radius=0.02];
		\draw[fill=black] (3+3*0.5/4,0) circle [radius=0.02];
		\node[] at (0,-0.3) {\small{$X_1$}};
		\node[] at (4,-0.3) {\small{$X_n$}};
		\node[] at (1.5-0.25,-0.3) {\small{$\tilde{X}_{i-1}$}};
		\node[] at (2.5+0.25,-0.3) {\small{$\tilde{X}_{i+1}$}};
		\node[above] at (2,0.01) {\small{$\tilde{Y}_{i{-}1,i{+}1}$}};
	\end{tikzpicture}
\end{equation}
for $1 < i < n$. The equations~(\ref{eq:tildeXY}) and~(\ref{eq:XXtoq}) can be recast as
\begin{equation}
\begin{aligned}
    \tilde{X}_{i-1}+\tilde{Y}_{i-1,i+1}&=X_{i-1}+Y_{i-1,i}\,,&\qquad \tilde{X}_{i+1}+\tilde{Y}_{i-1,i+1}&=X_{i+1}+Y_{i,i+1}\,,\\
    \tilde{X}_{i-1}-\tilde{Y}_{i-1,i+1}&=X_{i-1}+X_i-Y_{i,i+1}\,,&\qquad\tilde{X}_{i+1}-\tilde{Y}_{i-1,i+1}&=X_{i}+X_{i+1}-Y_{i-1,i}\,,\\
    \tilde{X}_{i-1}+\tilde{X}_{i+1}&=X_{i-1}+X_{i}+X_{i+1}\,.
\end{aligned}
\end{equation}
We construct the $C_{n-1}$ matrix associated to the shifted kinematics as
\begin{equation}
\resizebox{0.935\textwidth}{!}{$
\begin{aligned}
&C_{n-1}(\ldots,\tilde{X}_{i-1},\tilde{X}_{i+1},\ldots;\ldots,\tilde{Y}_{i-1,i+1},\ldots)\\
&=\begin{pmatrix}
    1&\cdots &1 &1 &1& \cdots &0\\
    0&\cdots &X_{1,i-2} {+} \tilde{X}_{i-1}{+}\tilde{Y}_{i-1,i+1} &X_{1,i-2} {+} \tilde{X}_{i-1}{-}\tilde{Y}_{i-1,i+1} & X_{1,i-2} {+} \tilde{X}_{i-1}{+} \tilde{X}_{i+1}{+}Y_{i+1,i+2}& \cdots &1
\end{pmatrix}\\
&=\begin{pmatrix}
    1&\cdots &1 &1 &1& \cdots &0\\
    0&\cdots & X_{1,i-1}^+ & X_{1,i}^- & X_{1,i+1}^+& \cdots &1
\end{pmatrix}
=\begin{pmatrix}
    1&\cdots &1 &1&1&\cdots &0\\
    z_1&\cdots &z_{2i-2}&z_{2i+1}&z_{2i+2}&\cdots &1
    \end{pmatrix},
\end{aligned}
$}
\end{equation}
where we used $X_{1,i-2} = \sum_{j=1}^{i-2}X_j$.
We see that this recovers exactly~\eqref{eq:Cpn}, but with the columns containing the two points $z_{2i-1} = X_{1,i-1}^-$ and $z_{2i} = X_{1,i}^+$ removed.
We conclude that this term is simply
\begin{equation}
    \psi_{n-1}(\ldots,\tilde{X}_{i-1},\tilde{X}_{i+1},\ldots;\ldots,\tilde{Y}_{i-1,i+1},\ldots) = \psi_{n-1}(1,\ldots,\widehat{2i{-}1},\widehat{2i},\ldots,2n)\,.
\end{equation}

The same dropout of two adjacent $z$'s occurs for the shifted kinematics associated to each of the terms in~(\ref{eq:dchain}), which with the help of~(\ref{eq:XXtoq}) can then be written as
\begin{equation}\label{eq:diffqp}
\begin{aligned}
    &\Delta_{n-1,1}\psi_{n}=
        (\psi_{n-1}({\hat{1},\hat{2}})-\psi_{n-1}(\hat{2},\hat{3}))\otimes\log q_{1,3,2n+1,2}\\
        +&\sum_{i=2}^{n-1}\Big(\psi_{n-1}(\widehat{2i{-}1},\widehat{2i}){\otimes}\log q_{2i-1,2i{+}1,2i-2,2i}\\
        &-\psi_{n-1}(\widehat{2i},\widehat{2i{+}1}){\otimes}\log q_{2i-1,2i{+}1,2n+1,2i}- \psi_{n-1}(\widehat{2i{-}2},\widehat{2i{-}1})){\otimes}\log q_{2i-2,2i,2i-1,2n+1}\Big)\\
        {+}&(\psi_{n-1}({\widehat{2n{-}1},\widehat{2n})}-\psi_{n-1}(\widehat{2n{-}2},\widehat{2n{-}1}))\otimes \log q_{2n-2,2n,2n-1,2n+1}\,.
\end{aligned}
\end{equation}
Collecting terms with the same wavefunction coefficients and using the identities $q_{ijkl} = q_{klij}$ and $q_{2 i-1,2 i+1,2 n+1,2 i}\, q_{2 i,2 i+2,2 i+1,2 n+1}=q_{2 i-1,2 i+1,2 i+2,2 i}$ allows this to be written as
\begin{equation}
\begin{aligned}
    \Delta_{n-1,1}\psi_{n} &=
   \sum_{i=0}^{n-1}  \psi_{n-1}(\widehat{2i{+}1}, \widehat{2i{+}2})\otimes \log q_{2i,2i{+}2,2i{+}1,2i+3}\\
   &-\sum_{i=1}^{n-1}   \psi_{n-1}(\widehat{2i}, \widehat{2i{+}1}) \otimes \log q_{2i-1,2i{+}1,2i{+}2,2i}\,,
\label{eq:simpd}
\end{aligned}
\end{equation}
which combines nicely into
\begin{equation}\label{eq:simplifiedDelta}
\boxed{
    \Delta_{n-1,1}\psi_{n}
    = \sum_{i=1}^{2n-1}   \psi_{n-1}(\widehat{i}, \widehat{i{+}1}) \otimes \qli_1^{(-)^{i-1}}( i{-}1,i,i{+}1,i{+}2)\,, \qquad n \ge 2\,,}
\end{equation}
in terms of the quadrangular polylogarithm~(\ref{eq:QLiquad}), where all indices are taken modulo $2n{+}1$. We emphasize that this equation is merely a rewriting of~(\ref{eq:dchain}), although the greatly simplified form attests to the utility of the parameterization~(\ref{eq:Cpn}).  In the next section we solve this recursion explicitly in terms of QLi functions.

\section{Solution of the Recursion}
\label{sec:solution}
\subsection{Main Formula}
\label{sec:mainformula}

Our main result is that the $n$-site chain wavefunction coefficient in dS space is given for $n \ge 1$ by the formula
\begin{equation}\label{eq:generalP}
\boxed{
\psi_n = F_n(2n{+}1, 1,\ldots, 2n{-}1) - F_n(2n,1,\ldots,2n{-}1) + F_n(2n,2n{+}1,2,\ldots,2n{-}1)\,.
}
\end{equation}
We consider the special case $n=1$ in the next section. For $n\geq 2$, the function $F_n$ has the universal form
\begin{equation}
\label{eq:H_function_final}
\boxed{
\begin{aligned}
       & F_n(0,\ldots,2n{-}1) = \\
        &\sum_{k=0}^{n-2} \sum_{ \substack{ S_0 \sqcup S_1 \sqcup \cdots \sqcup S_k \\ = \{0,\ldots, 2n-1\} } }\hspace{-1em} (-1)^{k+n} \left( \QLi^+_{\frac{|S_0|}{2}}(S_0) + \QLi^+_{\frac{|S_0|}{2}-1}(S_0) \log(q_{S_0})  \right) \prod_{i=1}^{k} \QLi^{(-)^{S_i(1)}}_{\frac{|S_i|}{2}-1}(S_i) \,,
\end{aligned}
}
\end{equation}
where $S_i(j)$ is the $j$-th element of $S_i$.  Here the second sum runs over all dissections of the polygon $(0,\ldots,2n{-}1)$ into even sub-polygons $\{S_0,\ldots,S_k\}$, with $S_0$ containing the root edge $(0,2n{-}1)$, to which we associate the cross-ratio given by
\begin{equation}
q_{S_0} = \prod_{i=1}^{\frac{|S_0|}{2}-1} q_{0,S_0(2i),S_0(2i+1),S_0(2i+2)}\,.
\end{equation}
The function $F$, by construction, satisfies cyclicity under a shift by two units:
\begin{align}
    F_n(2,3,\ldots,2n{-}1,0,1) = F_n(0,1,\ldots,2n{-}1)\,.
\end{align}
Its behavior under shifting by one is considerably more complicated; see~\eqref{eq:Fidentity}.

In Section~\ref{sec:examples} we demonstrate that~(\ref{eq:generalP}) reproduces the new expressions obtained for the two- and three-site chains in Sections~\ref{sec:prelude2} and~\ref{sec:prelude3}.  Then in Section~\ref{sec:proof} we provide a recursive proof that the total differential of~(\ref{eq:generalP}) (equivalently, the $(n{-}1,1)$ coproduct component) matches the known answer~\cite{He:2024olr} for all $n$, and also show that it vanishes in any soft limit $Y_{i,i+1} \to 0$, so it is not missing any additive numerical constant. Together, these steps constitute a proof of~(\ref{eq:generalP}) for all $n$.

\subsection{Examples}
\label{sec:examples}

We start with $n=1$ which we did not previously consider because it is somewhat of a special case. However, it still can be written in the form~\eqref{eq:generalP} in terms of
\begin{equation}
    F_1(a,b) = -\log \Delta_{ab}\,,
\end{equation}
where $\Delta_{ab}$ denotes the minor formed from columns $a, b$ of the $G(2,3)$ matrix
\begin{equation}
    C_1=
    \begin{pmatrix}
        1&1&0\\
        0&X_1&1
    \end{pmatrix}.
\end{equation}
The single-point wavefunction is then
\begin{equation}
\begin{aligned}
\psi_1&=F_1(3,1)-F_1(2,1)+F_1(2,3)\\
&=-\log\Delta_{31}+\log\Delta_{21}-\log\Delta_{23}\\
&=-\log{(-1)}+\log{(-X_1)}-\log{(1)}\\
&=\log{(X_1)}
\end{aligned}
\end{equation}
which is the correct physical result.

For $n=2$ there is just one quadrangulation so the sum in~\eqref{eq:H_function_final} has only one term, namely $k=0$, $S_0=(0,1,2,3)$, with $|S_0|=4$, and therefore gives
\begin{align}
\label{eq:Ffour}
    F_2(0,1,2,3)=\QLi^+_2(0,1,2,3)+\QLi^+_1(0,1,2,3)\log(q_{0123})\,,
\end{align}
which is identical to~(\ref{eq:Fquad}) after recalling~(\ref{eq:QLiquad}).

For $n=3$ the formula~(\ref{eq:H_function_final}) produces a total of 4 terms.  For $k=0$ there is a single term, corresponding to the hexagon itself.  For $k=1$ there are 3 terms, corresponding to the three ways of quadrangulating the hexagon. Altogether we have
\begin{equation}
\begin{aligned}
    F_3(0,1,2,3,4,5)=&-(\qli^+_3(0,1,2,3,4,5)+\qli^+_2(0,1,2,3,4,5) \log(q_{012345}))\\
    &+(\qli^+_2(0,3,4,5)+\qli^+_1(0,3,4,5)\log(q_{0345}))\qli^+_1(0,1,2,3)\\
    &+(\qli^+_2(0,1,4,5)+\qli^+_1(0,1,4,5)\log(q_{0145}))\qli^-_1(1,2,3,4)\\
    &+(\qli^+_2(0,1,2,5)+\qli^+_1(0,1,2,5)\log(q_{0125}))\qli^+_1(2,3,4,5)\,,
\end{aligned}
\end{equation}
which correspond respectively to the four terms in~(\ref{eq:hexagonF}) after using~(\ref{eq:Ffour}).

For $n=4$ the formula~(\ref{eq:H_function_final}) produces a total of 21 terms.
For $k=0$ there is a single term, the octagon itself.  For $k=1$ there are 8 terms, corresponding to the eight ways of dividing the octagon into a hexagon and a quadrilateral; 3 of these have the root in the quadrilateral and 5 have the root in the hexagon.  Finally for $k=2$ there are 12 terms, corresponding to the twelve ways of quadrangulating the octagon.  Altogether we have
\begin{equation}
\begin{aligned}
    F_4(0,1,2,3,4,5,6,7) =
\color{c1}\begin{tikzpicture}[baseline=(current bounding box.center), scale=0.5]
\foreach \i in {1,...,8} {
    \coordinate (V\i) at ({cos(90 - \i*45 + 22.5)}, {sin(90 - \i*45 + 22.5)});
}
\foreach \i in {1,...,8} {
    \pgfmathtruncatemacro{\j}{mod(\i,8)+1}
    \draw (V\i) -- (V\j);
}
\foreach \i in {1,...,8} {
    \pgfmathtruncatemacro{\j}{\i-1}
    \node at ({1.25*cos(90 - (\i+4)*45 + 22.5)}, {1.25*sin(90 - (\i+4)*45 + 22.5)}) {\tiny \(\j\)};
}
\draw (V4) -- (V5) node[midway] {\tiny $\times$};
\end{tikzpicture}\color{black}
-
\color{c2}\begin{tikzpicture}[baseline=(current bounding box.center), scale=0.5]
\foreach \i in {1,...,8} {
    \coordinate (V\i) at ({cos(90 - \i*45 + 22.5)}, {sin(90 - \i*45 + 22.5)});
}
\foreach \i in {1,...,8} {
    \pgfmathtruncatemacro{\j}{mod(\i,8)+1}
    \draw (V\i) -- (V\j);
}
\foreach \i in {1,...,8} {
    \pgfmathtruncatemacro{\j}{\i-1}
    \node at ({1.25*cos(90 - (\i+4)*45 + 22.5)}, {1.25*sin(90 - (\i+4)*45 + 22.5)}) {\tiny \(\j\)};
}
\draw (V4) -- (V5) node[midway] {\tiny $\times$};
\draw (V2) -- (V5);
\end{tikzpicture}\color{black}
-
\color{c2}\begin{tikzpicture}[baseline=(current bounding box.center), scale=0.5]
\foreach \i in {1,...,8} {
    \coordinate (V\i) at ({cos(90 - \i*45 + 22.5)}, {sin(90 - \i*45 + 22.5)});
}
\foreach \i in {1,...,8} {
    \pgfmathtruncatemacro{\j}{mod(\i,8)+1}
    \draw (V\i) -- (V\j);
}
\foreach \i in {1,...,8} {
    \pgfmathtruncatemacro{\j}{\i-1}
    \node at ({1.25*cos(90 - (\i+4)*45 + 22.5)}, {1.25*sin(90 - (\i+4)*45 + 22.5)}) {\tiny \(\j\)};
}
\draw (V4) -- (V5) node[midway] {\tiny $\times$};
\draw (V3) -- (V6);
\end{tikzpicture}\color{black}
&-
\color{c2}\begin{tikzpicture}[baseline=(current bounding box.center), scale=0.5]
\foreach \i in {1,...,8} {
    \coordinate (V\i) at ({cos(90 - \i*45 + 22.5)}, {sin(90 - \i*45 + 22.5)});
}
\foreach \i in {1,...,8} {
    \pgfmathtruncatemacro{\j}{mod(\i,8)+1}
    \draw (V\i) -- (V\j);
}
\foreach \i in {1,...,8} {
    \pgfmathtruncatemacro{\j}{\i-1}
    \node at ({1.25*cos(90 - (\i+4)*45 + 22.5)}, {1.25*sin(90 - (\i+4)*45 + 22.5)}) {\tiny \(\j\)};
}
\draw (V4) -- (V5) node[midway] {\tiny $\times$};
\draw (V4) -- (V7);
\end{tikzpicture}\color{black}
\\
-\color{c3}\begin{tikzpicture}[baseline=(current bounding box.center), scale=0.5]
\foreach \i in {1,...,8} {
    \coordinate (V\i) at ({cos(90 - \i*45 + 22.5)}, {sin(90 - \i*45 + 22.5)});
}
\foreach \i in {1,...,8} {
    \pgfmathtruncatemacro{\j}{mod(\i,8)+1}
    \draw (V\i) -- (V\j);
}
\foreach \i in {1,...,8} {
    \pgfmathtruncatemacro{\j}{\i-1}
    \node at ({1.25*cos(90 - (\i+4)*45 + 22.5)}, {1.25*sin(90 - (\i+4)*45 + 22.5)}) {\tiny \(\j\)};
}
\draw (V4) -- (V5) node[midway] {\tiny $\times$};
\draw (V5) -- (V8);
\end{tikzpicture}\color{black}
-
\color{c3}\begin{tikzpicture}[baseline=(current bounding box.center), scale=0.5]
\foreach \i in {1,...,8} {
    \coordinate (V\i) at ({cos(90 - \i*45 + 22.5)}, {sin(90 - \i*45 + 22.5)});
}
\foreach \i in {1,...,8} {
    \pgfmathtruncatemacro{\j}{mod(\i,8)+1}
    \draw (V\i) -- (V\j);
}
\foreach \i in {1,...,8} {
    \pgfmathtruncatemacro{\j}{\i-1}
    \node at ({1.25*cos(90 - (\i+4)*45 + 22.5)}, {1.25*sin(90 - (\i+4)*45 + 22.5)}) {\tiny \(\j\)};
}
\draw (V4) -- (V5) node[midway] {\tiny $\times$};
\draw (V6) -- (V1);
\end{tikzpicture}\color{black}
-
\color{c3}\begin{tikzpicture}[baseline=(current bounding box.center), scale=0.5]
\foreach \i in {1,...,8} {
    \coordinate (V\i) at ({cos(90 - \i*45 + 22.5)}, {sin(90 - \i*45 + 22.5)});
}
\foreach \i in {1,...,8} {
    \pgfmathtruncatemacro{\j}{mod(\i,8)+1}
    \draw (V\i) -- (V\j);
}
\foreach \i in {1,...,8} {
    \pgfmathtruncatemacro{\j}{\i-1}
    \node at ({1.25*cos(90 - (\i+4)*45 + 22.5)}, {1.25*sin(90 - (\i+4)*45 + 22.5)}) {\tiny \(\j\)};
}
\draw (V4) -- (V5) node[midway] {\tiny $\times$};
\draw (V7) -- (V2);
\end{tikzpicture}\color{black}
-
\color{c3}\begin{tikzpicture}[baseline=(current bounding box.center), scale=0.5]
\foreach \i in {1,...,8} {
    \coordinate (V\i) at ({cos(90 - \i*45 + 22.5)}, {sin(90 - \i*45 + 22.5)});
}
\foreach \i in {1,...,8} {
    \pgfmathtruncatemacro{\j}{mod(\i,8)+1}
    \draw (V\i) -- (V\j);
}
\foreach \i in {1,...,8} {
    \pgfmathtruncatemacro{\j}{\i-1}
    \node at ({1.25*cos(90 - (\i+4)*45 + 22.5)}, {1.25*sin(90 - (\i+4)*45 + 22.5)}) {\tiny \(\j\)};
}
\draw (V4) -- (V5) node[midway] {\tiny $\times$};
\draw (V8) -- (V3);
\end{tikzpicture}\color{black}
&-
\color{c3}\begin{tikzpicture}[baseline=(current bounding box.center), scale=0.5]
\foreach \i in {1,...,8} {
    \coordinate (V\i) at ({cos(90 - \i*45 + 22.5)}, {sin(90 - \i*45 + 22.5)});
}
\foreach \i in {1,...,8} {
    \pgfmathtruncatemacro{\j}{mod(\i,8)+1}
    \draw (V\i) -- (V\j);
}
\foreach \i in {1,...,8} {
    \pgfmathtruncatemacro{\j}{\i-1}
    \node at ({1.25*cos(90 - (\i+4)*45 + 22.5)}, {1.25*sin(90 - (\i+4)*45 + 22.5)}) {\tiny \(\j\)};
}
\draw (V4) -- (V5) node[midway] {\tiny $\times$};
\draw (V1) -- (V4);
\end{tikzpicture}\color{black}
\\
+
\color{c4}\begin{tikzpicture}[baseline=(current bounding box.center), scale=0.5]
\foreach \i in {1,...,8} {
    \coordinate (V\i) at ({cos(90 - \i*45 + 22.5)}, {sin(90 - \i*45 + 22.5)});
}
\foreach \i in {1,...,8} {
    \pgfmathtruncatemacro{\j}{mod(\i,8)+1}
    \draw (V\i) -- (V\j);
}
\foreach \i in {1,...,8} {
    \pgfmathtruncatemacro{\j}{\i-1}
    \node at ({1.25*cos(90 - (\i+4)*45 + 22.5)}, {1.25*sin(90 - (\i+4)*45 + 22.5)}) {\tiny \(\j\)};
}
\draw (V4) -- (V5) node[midway] {\tiny $\times$};
\draw (V8) -- (V3);
\draw (V5) -- (V8);
\end{tikzpicture}\color{black}
+
\color{c4}\begin{tikzpicture}[baseline=(current bounding box.center), scale=0.5]
\foreach \i in {1,...,8} {
    \coordinate (V\i) at ({cos(90 - \i*45 + 22.5)}, {sin(90 - \i*45 + 22.5)});
}
\foreach \i in {1,...,8} {
    \pgfmathtruncatemacro{\j}{mod(\i,8)+1}
    \draw (V\i) -- (V\j);
}
\foreach \i in {1,...,8} {
    \pgfmathtruncatemacro{\j}{\i-1}
    \node at ({1.25*cos(90 - (\i+4)*45 + 22.5)}, {1.25*sin(90 - (\i+4)*45 + 22.5)}) {\tiny \(\j\)};
}
\draw (V4) -- (V5) node[midway] {\tiny $\times$};
\draw (V1) -- (V4);
\draw (V5) -- (V8);
\end{tikzpicture}\color{black}
+
\color{c4}\begin{tikzpicture}[baseline=(current bounding box.center), scale=0.5]
\foreach \i in {1,...,8} {
    \coordinate (V\i) at ({cos(90 - \i*45 + 22.5)}, {sin(90 - \i*45 + 22.5)});
}
\foreach \i in {1,...,8} {
    \pgfmathtruncatemacro{\j}{mod(\i,8)+1}
    \draw (V\i) -- (V\j);
}
\foreach \i in {1,...,8} {
    \pgfmathtruncatemacro{\j}{\i-1}
    \node at ({1.25*cos(90 - (\i+4)*45 + 22.5)}, {1.25*sin(90 - (\i+4)*45 + 22.5)}) {\tiny \(\j\)};
}
\draw (V4) -- (V5) node[midway] {\tiny $\times$};
\draw (V6) -- (V1);
\draw (V1) -- (V4);
\end{tikzpicture}\color{black}
+
\color{c4}\begin{tikzpicture}[baseline=(current bounding box.center), scale=0.5]
\foreach \i in {1,...,8} {
    \coordinate (V\i) at ({cos(90 - \i*45 + 22.5)}, {sin(90 - \i*45 + 22.5)});
}
\foreach \i in {1,...,8} {
    \pgfmathtruncatemacro{\j}{mod(\i,8)+1}
    \draw (V\i) -- (V\j);
}
\foreach \i in {1,...,8} {
    \pgfmathtruncatemacro{\j}{\i-1}
    \node at ({1.25*cos(90 - (\i+4)*45 + 22.5)}, {1.25*sin(90 - (\i+4)*45 + 22.5)}) {\tiny \(\j\)};
}
\draw (V4) -- (V5) node[midway] {\tiny $\times$};
\draw (V5) -- (V8);
\draw (V5) -- (V2);
\end{tikzpicture}\color{black}
&+
\color{c4}\begin{tikzpicture}[baseline=(current bounding box.center), scale=0.5]
\foreach \i in {1,...,8} {
    \coordinate (V\i) at ({cos(90 - \i*45 + 22.5)}, {sin(90 - \i*45 + 22.5)});
}
\foreach \i in {1,...,8} {
    \pgfmathtruncatemacro{\j}{mod(\i,8)+1}
    \draw (V\i) -- (V\j);
}
\foreach \i in {1,...,8} {
    \pgfmathtruncatemacro{\j}{\i-1}
    \node at ({1.25*cos(90 - (\i+4)*45 + 22.5)}, {1.25*sin(90 - (\i+4)*45 + 22.5)}) {\tiny \(\j\)};
}
\draw (V4) -- (V5) node[midway] {\tiny $\times$};
\draw (V5) -- (V2);
\draw (V6) -- (V1);
\end{tikzpicture}\color{black}
+
\color{c4}\begin{tikzpicture}[baseline=(current bounding box.center), scale=0.5]
\foreach \i in {1,...,8} {
    \coordinate (V\i) at ({cos(90 - \i*45 + 22.5)}, {sin(90 - \i*45 + 22.5)});
}
\foreach \i in {1,...,8} {
    \pgfmathtruncatemacro{\j}{mod(\i,8)+1}
    \draw (V\i) -- (V\j);
}
\foreach \i in {1,...,8} {
    \pgfmathtruncatemacro{\j}{\i-1}
    \node at ({1.25*cos(90 - (\i+4)*45 + 22.5)}, {1.25*sin(90 - (\i+4)*45 + 22.5)}) {\tiny \(\j\)};
}
\draw (V4) -- (V5) node[midway] {\tiny $\times$};
\draw (V5) -- (V2);
\draw (V7) -- (V2);
\end{tikzpicture}\color{black}
\\
+
\color{c4}\begin{tikzpicture}[baseline=(current bounding box.center), scale=0.5]
\foreach \i in {1,...,8} {
    \coordinate (V\i) at ({cos(90 - \i*45 + 22.5)}, {sin(90 - \i*45 + 22.5)});
}
\foreach \i in {1,...,8} {
    \pgfmathtruncatemacro{\j}{mod(\i,8)+1}
    \draw (V\i) -- (V\j);
}
\foreach \i in {1,...,8} {
    \pgfmathtruncatemacro{\j}{\i-1}
    \node at ({1.25*cos(90 - (\i+4)*45 + 22.5)}, {1.25*sin(90 - (\i+4)*45 + 22.5)}) {\tiny \(\j\)};
}
\draw (V4) -- (V5) node[midway] {\tiny $\times$};
\draw (V6) -- (V1);
\draw (V6) -- (V3);
\end{tikzpicture}\color{black}
+
\color{c4}\begin{tikzpicture}[baseline=(current bounding box.center), scale=0.5]
\foreach \i in {1,...,8} {
    \coordinate (V\i) at ({cos(90 - \i*45 + 22.5)}, {sin(90 - \i*45 + 22.5)});
}
\foreach \i in {1,...,8} {
    \pgfmathtruncatemacro{\j}{mod(\i,8)+1}
    \draw (V\i) -- (V\j);
}
\foreach \i in {1,...,8} {
    \pgfmathtruncatemacro{\j}{\i-1}
    \node at ({1.25*cos(90 - (\i+4)*45 + 22.5)}, {1.25*sin(90 - (\i+4)*45 + 22.5)}) {\tiny \(\j\)};
}
\draw (V4) -- (V5) node[midway] {\tiny $\times$};
\draw (V6) -- (V3);
\draw (V7) -- (V2);
\end{tikzpicture}\color{black}
+
\color{c4}\begin{tikzpicture}[baseline=(current bounding box.center), scale=0.5]
\foreach \i in {1,...,8} {
    \coordinate (V\i) at ({cos(90 - \i*45 + 22.5)}, {sin(90 - \i*45 + 22.5)});
}
\foreach \i in {1,...,8} {
    \pgfmathtruncatemacro{\j}{mod(\i,8)+1}
    \draw (V\i) -- (V\j);
}
\foreach \i in {1,...,8} {
    \pgfmathtruncatemacro{\j}{\i-1}
    \node at ({1.25*cos(90 - (\i+4)*45 + 22.5)}, {1.25*sin(90 - (\i+4)*45 + 22.5)}) {\tiny \(\j\)};
}
\draw (V4) -- (V5) node[midway] {\tiny $\times$};
\draw (V6) -- (V3);
\draw (V8) -- (V3);
\end{tikzpicture}\color{black}
+
\color{c4}\begin{tikzpicture}[baseline=(current bounding box.center), scale=0.5]
\foreach \i in {1,...,8} {
    \coordinate (V\i) at ({cos(90 - \i*45 + 22.5)}, {sin(90 - \i*45 + 22.5)});
}
\foreach \i in {1,...,8} {
    \pgfmathtruncatemacro{\j}{mod(\i,8)+1}
    \draw (V\i) -- (V\j);
}
\foreach \i in {1,...,8} {
    \pgfmathtruncatemacro{\j}{\i-1}
    \node at ({1.25*cos(90 - (\i+4)*45 + 22.5)}, {1.25*sin(90 - (\i+4)*45 + 22.5)}) {\tiny \(\j\)};
}
\draw (V4) -- (V5) node[midway] {\tiny $\times$};
\draw (V7) -- (V2);
\draw (V7) -- (V4);
\end{tikzpicture}\color{black}
&+
\color{c4}\begin{tikzpicture}[baseline=(current bounding box.center), scale=0.5]
\foreach \i in {1,...,8} {
    \coordinate (V\i) at ({cos(90 - \i*45 + 22.5)}, {sin(90 - \i*45 + 22.5)});
}
\foreach \i in {1,...,8} {
    \pgfmathtruncatemacro{\j}{mod(\i,8)+1}
    \draw (V\i) -- (V\j);
}
\foreach \i in {1,...,8} {
    \pgfmathtruncatemacro{\j}{\i-1}
    \node at ({1.25*cos(90 - (\i+4)*45 + 22.5)}, {1.25*sin(90 - (\i+4)*45 + 22.5)}) {\tiny \(\j\)};
}
\draw (V4) -- (V5) node[midway] {\tiny $\times$};
\draw (V7) -- (V4);
\draw (V8) -- (V3);
\end{tikzpicture}\color{black}
+
\color{c4}\begin{tikzpicture}[baseline=(current bounding box.center), scale=0.5]
\foreach \i in {1,...,8} {
    \coordinate (V\i) at ({cos(90 - \i*45 + 22.5)}, {sin(90 - \i*45 + 22.5)});
}
\foreach \i in {1,...,8} {
    \pgfmathtruncatemacro{\j}{mod(\i,8)+1}
    \draw (V\i) -- (V\j);
}
\foreach \i in {1,...,8} {
    \pgfmathtruncatemacro{\j}{\i-1}
    \node at ({1.25*cos(90 - (\i+4)*45 + 22.5)}, {1.25*sin(90 - (\i+4)*45 + 22.5)}) {\tiny \(\j\)};
}
\draw (V4) -- (V5) node[midway] {\tiny $\times$};
\draw (V7) -- (V4);
\draw (V1) -- (V4);
\end{tikzpicture}\color{black}
\end{aligned}
\end{equation}
where the four different types of terms are represented by
\addtocounter{equation}{1}
\begin{align*}
\color{c1}\begin{tikzpicture}[baseline=(current bounding box.center), scale=0.5]
\foreach \i in {1,...,8} {
    \coordinate (V\i) at ({cos(90 - \i*45 + 22.5)}, {sin(90 - \i*45 + 22.5)});
}
\foreach \i in {1,...,8} {
    \pgfmathtruncatemacro{\j}{mod(\i,8)+1}
    \draw (V\i) -- (V\j);
}
\foreach \i in {1,...,8} {
    \pgfmathtruncatemacro{\j}{\i-1}
    \node at ({1.25*cos(90 - (\i+4)*45 + 22.5)}, {1.25*sin(90 - (\i+4)*45 + 22.5)}) {\tiny \(\j\)};
}
\draw (V4) -- (V5) node[midway] {\tiny $\times$};
\end{tikzpicture}\color{black}
&= \QLi_4^+(0,1,2,3,4,5,6,7) + \QLi_3^+(0,1,2,3,4,5,6,7) \log(q_{01234567})\,,
\\
\color{c2}\begin{tikzpicture}[baseline=(current bounding box.center), scale=0.5]
\foreach \i in {1,...,8} {
    \coordinate (V\i) at ({cos(90 - \i*45 + 22.5)}, {sin(90 - \i*45 + 22.5)});
}
\foreach \i in {1,...,8} {
    \pgfmathtruncatemacro{\j}{mod(\i,8)+1}
    \draw (V\i) -- (V\j);
}
\foreach \i in {1,...,8} {
    \pgfmathtruncatemacro{\j}{\i-1}
    \node at ({1.25*cos(90 - (\i+4)*45 + 22.5)}, {1.25*sin(90 - (\i+4)*45 + 22.5)}) {\tiny \(\j\)};
}
\draw (V4) -- (V5) node[midway] {\tiny $\times$};
\draw (V2) -- (V5);
\end{tikzpicture}\color{black}
&= \left( \QLi_2^+(0,5,6,7) + \QLi_1^+(0,5,6,7) \log(q_{0567}) \right) \QLi_2^+(0,1,2,3,4,5)\,,
\tag*{\raisebox{-\baselineskip}{(\theequation)}}
\\
\color{c3}\begin{tikzpicture}[baseline=(current bounding box.center), scale=0.5]
\foreach \i in {1,...,8} {
    \coordinate (V\i) at ({cos(90 - \i*45 + 22.5)}, {sin(90 - \i*45 + 22.5)});
}
\foreach \i in {1,...,8} {
    \pgfmathtruncatemacro{\j}{mod(\i,8)+1}
    \draw (V\i) -- (V\j);
}
\foreach \i in {1,...,8} {
    \pgfmathtruncatemacro{\j}{\i-1}
    \node at ({1.25*cos(90 - (\i+4)*45 + 22.5)}, {1.25*sin(90 - (\i+4)*45 + 22.5)}) {\tiny \(\j\)};
}
\draw (V4) -- (V5) node[midway] {\tiny $\times$};
\draw (V6) -- (V1);
\end{tikzpicture}\color{black}
&= \left( \QLi_3^+(0,1,4,5,6,7) + \QLi_2^+(0,1,4,5,6,7) \log(q_{014567}) \right) \QLi^-_1(1,2,3,4)\,,
\\
\color{c4}\begin{tikzpicture}[baseline=(current bounding box.center), scale=0.5]
\foreach \i in {1,...,8} {
    \coordinate (V\i) at ({cos(90 - \i*45 + 22.5)}, {sin(90 - \i*45 + 22.5)});
}
\foreach \i in {1,...,8} {
    \pgfmathtruncatemacro{\j}{mod(\i,8)+1}
    \draw (V\i) -- (V\j);
}
\foreach \i in {1,...,8} {
    \pgfmathtruncatemacro{\j}{\i-1}
    \node at ({1.25*cos(90 - (\i+4)*45 + 22.5)}, {1.25*sin(90 - (\i+4)*45 + 22.5)}) {\tiny \(\j\)};
}
\draw (V4) -- (V5) node[midway] {\tiny $\times$};
\draw (V8) -- (V3);
\draw (V5) -- (V8);
\end{tikzpicture}\color{black}
&= \left( \QLi_2^+(0,3,6,7) + \QLi_1^+(0,3,6,7) \log (q_{0367})\right) \QLi_1^+(0,1,2,3) \QLi_1^-(3,4,5,6)\,.
\end{align*}

\subsection{Outline of the Proof}
\label{sec:proof}

Here we outline the proof of our main results \eqref{eq:generalP} and \eqref{eq:H_function_final}.  To start with, let us assume that we have a collection of functions $F_2,F_3,\ldots$ starting with
\begin{equation}
    F_2(0,1,2,3)=\QLi^+_2(0,1,2,3)+\QLi^+_1(0,1,2,3)\log(q_{0123})
\end{equation}
and satisfying for $n>2$ the recursion
\begin{equation}\label{eq:coproFphy}
    \Delta_{n-1,1} F_{n}(0,\ldots,2n{-}1)=\sum_{i=0}^{2n-1}   F_{n-1}(0,\ldots,\widehat{i}, \widehat{i{+}1},\ldots,2n{-}1) \otimes \qli_1^{(-)^{i-1}}(i{-}1,i,i{+}1,i{+}2)\,,
\end{equation}
which is identical in form to~\eqref{eq:simplifiedDelta} except that the sum includes an extra $i=0$ term.
If we take the alternating sum of both sides of~\eqref{eq:coproFphy} with arguments as indicated in~(\ref{eq:generalP}), then the three $i=0$ terms cancel out because the first entry in each will be the same -- specifically, $F_{n-1}(2,\ldots,2n{-}1)$ -- and the second entries combine to zero thanks to the functional identity
\begin{equation}
\qli^{-}_{1}(2n{-}1,2n{+}1,1,2)-\qli^{-}_{1}(2n{-}1,2n,1,2)+\qli^{-}_{1}(2n{-}1,2n,2n{+}1,2)=0\,.
\end{equation}
Therefore, the combination
\begin{equation}
\label{eq:combination}
    F_n(2n{+}1, 1,\ldots, 2n{-}1) - F_n(2n,1,\ldots,2n{-}1) + F_n(2n,2n{+}1,2,\ldots,2n{-}1)
\end{equation}
satisfies~\eqref{eq:simplifiedDelta} and hence agrees with the wavefunction coefficient $\psi_n$ up to an additive constant. If we further demand that each $F_n$ vanishes in every soft limit $Y_{i,i+1} \to 0$, the same will be true of the combination~\eqref{eq:combination} and we can conclude that it computes the correct physical wavefunction coefficient $\psi_n$.  To summarize, if $F_n$ is a collection of functions that satisfy~\eqref{eq:coproFphy} and vanish in soft limits, then~\eqref{eq:generalP} is correct.

The second step of our proof is to show that the unique solution to~\eqref{eq:coproFphy} that vanishes in soft limits is given recursively for $n\ge 2$ by
\begin{multline}\label{eq:Hrecursion3}
        (-1)^n F_n(0,\ldots,2n{-}1) = \qli^{+}_{n}(0,\ldots,2n{-}1) + \qli^{+}_{n-1}(0,\ldots,2n{-}1) \log\left(q_{0,\ldots,2n-1}\right)\\
        + \sum_{k=1}^{n-2} \sum_{ \substack{ S_0 \sqcup S_1 \sqcup \cdots \sqcup S_k \\ = \{0,\cdots,2n-1\}, \\ \text{dim}(S_0 \cap S_\bullet) = 1 } } (-1)^{1+|S_0|/2} F_{\frac{|S_0|}{2}}(S_0) \prod_{i=1}^{k} \qli^{(-)^{S_i(1)}}_{|S_i|/2-1}(S_i)\,. 
\end{multline}
Similar to~\eqref{eq:H_function_final}, here the second sum runs over all dissections of the polygon $(0,\ldots,2n{-}1)$ into even sub-polygons $\{S_0,S_1,\ldots,S_k\}$, with $S_0$ containing the root edge $(0,2n{-}1)$. The additional condition $\text{dim}(S_0 \cap S_\bullet) = 1$ requires that each $S_i$ with $i\geq 1$ shares an edge with $S_0$. The proof of this formula takes two steps. First we prove that it satisfies~\eqref{eq:coproFphy}, and secondly we rule out any missing additive constant by considering soft limits. The details of this proof are given in Appendix~\ref{sec:recursionproof}.

The final step of the proof is to show that~\eqref{eq:Hrecursion3} implies~\eqref{eq:H_function_final}. Here we only sketch the proof. The equation~\eqref{eq:Hrecursion3} expresses $F_n(0,\ldots,2n{-}1)$ as an inhomogeneous term on the first line  plus a sum over dissections in which other polygons $S_{i \ne 0}$ are glued onto the root polygon $S_0$. We then ``unwind''~\eqref{eq:Hrecursion3} by repeatedly plugging in the left-hand side for each $F_{n>2}$  appearing on the right-hand side. This process stops whenever we reach an $F_2$, which is given by the inhomogeneous term alone, and the result is a sum over \emph{all} dissections of $(0,\ldots,2n{-}1)$ into even sub-polygons, each contributing a product of $\qli^\pm$ factors. We have checked that the total coefficient of every dissection is exactly the one given in~\eqref{eq:H_function_final}, which completes the proof.

\section{Outlook}

In the physics literature on scattering amplitudes in SYM theory, a symbol is said to satisfy cluster adjacency if, in each term, every pair of adjacent symbol letters is compatible, which means they belong to a common cluster.  This property has played a crucial role in bootstrapping six- and seven-point amplitudes, where the relevant cluster algebras are respectively $A_3$ and $E_6$. Rudenko's quadrangular polylogarithm functions associated to a $(2n{+}2)$-gon provide bases for functions satisfying a vastly stronger condition that we have called total compatibility: all symbol letters appearing in any given term are mutually compatible with respect to the $A_{2n-2}$ cluster algebra.  Amplitudes in SYM theory do not satisfy this stronger condition, so it has been an open problem to determine whether quadrangular polylogarithm functions have any direct application to physics.

In~\cite{Capuano:2026pgq} it was shown that the $n$-site chain graph cosmological wavefunction coefficient $\psi_n$ in dS space satisfies total compatibility with respect to the $A_{2n-2}$ cluster algebras, which immediately implies that $\psi_n$ \emph{can} be expressed in terms of quadrangular polylogarithms, but a priori there is no reason to expect such a representation to be particularly nice. However, in this paper we provide an explicit all-$n$ formula for $\psi_n$ in just two lines~\eqref{eq:generalP} and \eqref{eq:H_function_final}. The key step in proving this formula is to rewrite the recursive differential equation derived in~\cite{He:2024olr} in terms of $A_{2n-2}$ cluster variables, which allows it to be put into the form~\eqref{eq:simplifiedDelta} which we are able to directly relate to the coproduct formula~\eqref{eq:DeltaQLi} satisfied by quadrangular polylogarithms.

Our work suggests several natural directions for future work that we will report in a series of upcoming papers. It is interesting to extend the analysis to wavefunction coefficients associated to loop graphs, which were shown in~\cite{Paranjape:2026htn} to satisfy cluster adjacency (and in fact, by the same argument as in~\cite{Capuano:2026pgq}, total compatibility) with respect to $B$-type cluster algebras, or even to more general graphs.  It was argued in~\cite{Capuano:2025ehm,Paranjape:2026htn} that the ``kinematic flow'' equations~\cite{Arkani-Hamed:2023kig} imply that total compatibility persists at arbitrary order in the $\epsilon$ expansion around dS space, and it is interesting to see whether quadrangular polylogarithm functions could be relevant for higher-order terms.
We have considered only massless fields (i.e., conformally coupled fields in dS space), but kinematic flow equations for massive fields have recently been written down in~\cite{Baumann:2026atn} and it is natural to wonder whether they manifest any connection to cluster algebras.   Finally, it is interesting to study whether the same mathematical structures that we have found survive at the level of correlators, which are known to be simpler than wavefunction coefficients~\cite{Arkani-Hamed:2025mce}. In particular, dS correlators in conformally coupled $\phi^3$ theory have been studied recently in~\cite{Chowdhury:2026dwm}.

\acknowledgments

We are grateful to Lance Dixon, James Drummond, Shruti Paranjape, Marcos Skowronek and Cristian Vergu for helpful discussions and collaboration on related work. LF, T{\L}, LR, MS, AV, and Y-QZ would like to thank the Erwin Schr\"odinger International Institute for Mathematics and Physics (ESI) at the University of Vienna (Austria) for the opportunity and financial support to participate in the Thematic Programme ``Amplitudes and Algebraic Geometry" in 2026, which was instrumental to our collaboration. This work was also supported in part by the Deutsche Forschungsgemeinschaft (DFG, German Research Foundation) Projektnummer 508889767/FOR5582 ``Modern Foundations of Scattering Amplitudes" (LF, Y-QZ), by the Royal Society via a Newton International Fellowship (LR), by the US Department of Energy under contract DE-SC0010010 Task F (MS, AV, H-CW), and by Simons Investigator Award \#376208 (AV).

\appendix
\section{Details of the Proof}
\label{sec:recursionproof}

In this appendix we prove the functional recursion relation \eqref{eq:Hrecursion3} by induction on $n$. 
To start with, at $n=2$, \eqref{eq:Hrecursion3} gives
\begin{equation}
    F_2(0,1,2,3) = \qli^{+}_2(0,1,2,3) + \qli^{+}_{1}(0,1,2,3) \log(q_{0123})\,,
\end{equation}
which is consistent with~\eqref{eq:Fquad} after recalling~\eqref{eq:QLiquad}.  Next we prove that~\eqref{eq:Hrecursion3} is true for any specific value of $n$ assuming it is true for all smaller values of $n$.  There are two parts to the proof: first in Appendix~\ref{sec:A1} we prove that the formula holds at the level of the coproduct component $\Delta_{n-1,1}$, which implies equality up to an additive constant, and then in Appendix~\ref{sec:A2} we use soft limits to argue that no additive constant is missed.

\subsection{Coproduct}
\label{sec:A1}

Let us start by taking the $\Delta_{n-1,1}$ coproduct on both sides of~\eqref{eq:Hrecursion3}. From \eqref{eq:DeltaQLi}--\eqref{eq:DeltaQLi0}, we note that there are two kinds of terms in the coproduct of the $\QLi$ functions appearing on the right-hand side: those in which the second entry is $\log q$ for some cross-ratio $q$, which arise from the first line in~\eqref{eq:DeltaQLi}, and those in which the second entry is of the form $\QLi_1^\pm \sim \pm \log(1 - q^{\pm 1})$, which arise from all other terms in~\eqref{eq:DeltaQLi}--\eqref{eq:DeltaQLi0}. Our first step is to show that all terms of the former type cancel out.  From the first line of~\eqref{eq:Hrecursion3} we get
\begin{equation}
\begin{split}
    \Delta_{n-1,1} \left( \qli^{+}_{n}(S) + \qli^{+}_{n-1}(S)  \log\left(q_S\right) \right)
    &= \Delta_{n-1,1} \qli^{+}_{n}(S) + \QLi_{n-1}^+(S) \otimes \log q_S\\
&\quad+ (\Delta_{n-2,1} \QLi_{n-1}^+(S)) \cdot ( \log q_S \otimes 1)\,.
\end{split}
\end{equation}
Now it is evident from~\eqref{eq:DeltaQLi0} that the second line does not have any terms like $\log q$ in the second entry, and in the first line we see from~\eqref{eq:DeltaQLi} that the second term cancels the contribution from the first term in which the second entry is $\log q_S$. Now in the second line of~\eqref{eq:Hrecursion3}, the weight of each $\QLi$ function is such that~\eqref{eq:DeltaQLi0} applies, so there are no terms with $\log q$ in the second entry.  Finally, when $\Delta_{n-1,1}$ hits any of the $F$ functions, all of which have weight lower than $n$, our inductive use of~\eqref{eq:coproFphy} shows that there are no such terms.  Altogether, we conclude that $\Delta_{n-1,1} F_n$ only has terms with $\QLi_1^{(-)^a}(a,b,c,d)$ in the second entry, and no terms with $\log q$ in the second entry.

Recall that we want to show that the $\Delta_{n-1,1}$ coproduct of the right-hand side of~\eqref{eq:Hrecursion3} agrees with what one would obtain by using~\eqref{eq:coproFphy} on the left-hand side.  Instead of manipulating terms in the former to check that they can be rearranged to the latter, we find it more convenient to move the left-hand side to the right-hand side and then check that the sum of all terms can be rearranged to zero.  In this manner we arrive at the following formula that we aim to prove in this section:
\begin{equation}\label{eq:Hrecursion4} 
   \begin{aligned}
        0 = \Delta_{n-1,1} & \Bigg\lbrack \qli^{+}_{n}(S) + \qli^{+}_{n-1}(S)  \log\left(q_S\right)\\
        & + \sum_{k=0}^{n-2} \sum_{ \substack{ S_0 \sqcup S_1 \sqcup \cdots \sqcup S_k \\ = \{0,\cdots,2n-1\}, \\ \text{dim}(S_0 \cap S_\bullet) = 1 } } (-1)^{1+|S_0|/2} F_{\frac{|S_0|}{2}}(S_0) \prod_{i=1}^{k} \qli^{(-)^{S_i(1)}}_{|S_i|/2-1}(S_i) \Bigg\rbrack \,. \\
    \end{aligned}
\end{equation}
As we have argued, the coproduct of every surviving term in~\eqref{eq:Hrecursion4} has the form $\cdots\otimes\qli^{(-)^a}_1(a,b,c,d)$ where the argument $(a,b,c,d)$ is an alternating quadrangle.
Based on the position of this quadrangle within the original polygon $S$, we divide the analysis into two cases, depending on whether $(a,b,c,d)$ contains the root edge $(0,2n{-}1)$.

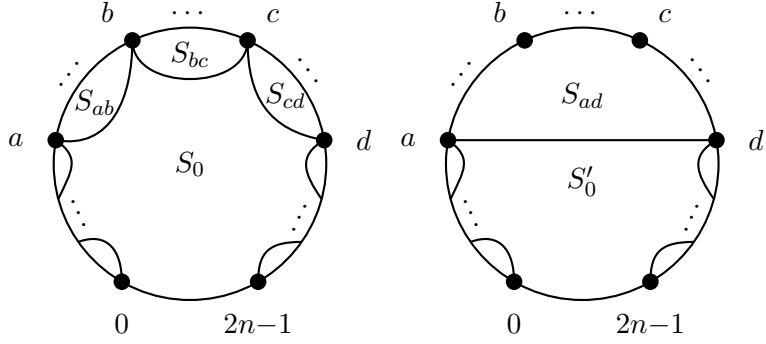
\begin{figure}
    \centering
    \begin{tikzpicture}
    [scale = 0.45,
    thick,
    dot/.style={circle,fill=black,inner sep=2.2pt},
    ]
    \coordinate (O) at (0,0);
    \def\R{4}
    \coordinate (a) at ($(O)+(170:\R)$);
    \coordinate (b) at ($(O)+(115:\R)$);
    \coordinate (c) at ($(O)+(65:\R)$);
    \coordinate (d) at ($(O)+(10:\R)$);
    \coordinate (z) at ($(O)+(240:\R)$);   
    \coordinate (n) at ($(O)+(300:\R)$);   
    \coordinate (al) at ($(O)+(195:\R)$);
    \coordinate (zl) at ($(O)+(215:\R)$);
    \coordinate (dr) at ($(O)+(-15:\R)$);
    \coordinate (nr) at ($(O)+(325:\R)$);
    
    \draw (O) circle (\R);
    
    \draw (a) to[out=-10,in=270,looseness=1.15] (b);
    \draw (b) to[out=270,in=270,looseness=1.15] (c);
    \draw (c) to[out=270,in=170,looseness=1.15] (d);
    
    \draw (a)  to[out=-35,in=65,looseness=1.35] (al);
    \draw (zl) to[out=20,in=90,looseness=1.35] (z);
    \draw (d)  to[out=-150,in=125,looseness=1.35] (dr);
    \draw (n)  to[out=90,in=-180,looseness=1.35] (nr);
    
    \node[dot] at (a) {};
    \node[dot] at (b) {};
    \node[dot] at (c) {};
    \node[dot] at (d) {};
    \node[dot] at (z) {};
    \node[dot] at (n) {};
    
    \node[left=8pt]  at (a) {$a$};
    \node[above left=3pt]  at (b) {$b$};
    \node[above right=3pt] at (c) {$c$};
    \node[right=8pt] at (d) {$d$};
    \node[below=8pt] at (z) {$0$};
    \node[below=8pt] at (n) {$2n{-}1$};

    \node[rotate= 52.5] at ($(O)+(142.5:1.10*\R)$) {$\cdots$};
    \node[rotate=  0  ] at ($(O)+( 90  :1.10*\R)$) {$\cdots$};
    \node[rotate=-52.5] at ($(O)+( 37.5:1.10*\R)$) {$\cdots$};
    \node[rotate=-65] at ($(O)+( 205:0.90*\R)$) {$\cdots$};
    \node[rotate=65] at ($(O)+( 335:0.90*\R)$) {$\cdots$};
    
    \node at (-2.8, 1.80) {$S_{ab}$};
    \node at ( 0.00, 3.2) {$S_{bc}$};
    \node at ( 2.8, 2) {$S_{cd}$};
    \node at ( 0.00, 0.00) {$S_0$};
    
    \end{tikzpicture}
    \begin{tikzpicture}[scale = 0.45,
    thick,
    dot/.style={circle,fill=black,inner sep=2.2pt},
    ]
    \coordinate (O) at (0,0);
    \def\R{4}
    \coordinate (a) at ($(O)+(170:\R)$);
    \coordinate (b) at ($(O)+(115:\R)$);
    \coordinate (c) at ($(O)+(65:\R)$);
    \coordinate (d) at ($(O)+(10:\R)$);
    \coordinate (z) at ($(O)+(240:\R)$);   
    \coordinate (n) at ($(O)+(300:\R)$);   
    \coordinate (al) at ($(O)+(195:\R)$);
    \coordinate (zl) at ($(O)+(215:\R)$);
    \coordinate (dr) at ($(O)+(-15:\R)$);
    \coordinate (nr) at ($(O)+(325:\R)$);
    
    \draw (O) circle (\R);
    
    \draw (a) to (d);
    
    \draw (a)  to[out=-35,in=65,looseness=1.35] (al);
    \draw (zl) to[out=20,in=90,looseness=1.35] (z);
    \draw (d)  to[out=-150,in=125,looseness=1.35] (dr);
    \draw (n)  to[out=90,in=-180,looseness=1.35] (nr);
    
    \node[dot] at (a) {};
    \node[dot] at (b) {};
    \node[dot] at (c) {};
    \node[dot] at (d) {};
    \node[dot] at (z) {};
    \node[dot] at (n) {};
    
    \node[left=8pt]  at (a) {$a$};
    \node[above left=3pt]  at (b) {$b$};
    \node[above right=3pt] at (c) {$c$};
    \node[right=8pt] at (d) {$d$};
    \node[below=8pt] at (z) {$0$};
    \node[below=8pt] at (n) {$2n{-}1$};

    \node[rotate= 52.5] at ($(O)+(142.5:1.10*\R)$) {$\cdots$};
    \node[rotate=  0  ] at ($(O)+( 90  :1.10*\R)$) {$\cdots$};
    \node[rotate=-52.5] at ($(O)+( 37.5:1.10*\R)$) {$\cdots$};
    \node[rotate=-65] at ($(O)+( 205:0.90*\R)$) {$\cdots$};
    \node[rotate=65] at ($(O)+( 335:0.90*\R)$) {$\cdots$};
    
    \node at ( 0.00, 2) {$S_{ad}$};
    \node at ( 0.00, -0.50) {$S'_0$};
    
    \end{tikzpicture}
    \caption{The two dissections that contribute to the $\Delta_{n-1,1}$ coproduct component of the form $\cdots\otimes\qli_1^{(-)^a}(a,b,c,d)$ for case (1) of~\eqref{eq:Hrecursion4}. On the left, we have $(a,b,c,d)$ on the boundary of $S_0$, and on the right, we have $(a,b,c,d)$ as a root quadrangle of $S_{ad}$.}
    \label{fig: case 1 proof}
\end{figure}

\paragraph{Case (1): non-rooted.}
The terms of the form $\cdots\otimes\qli_1^{(-)^a}(a,b,c,d)$ arise in two ways: either $(a,b,c,d)$ is a boundary quadrangle of $S_0$, or $(a,b,c,d)$ is the root quadrangle of some $S_l\equiv S_{ad}$, both shown in Figure \ref{fig: case 1 proof}. 
Following \eqref{eq:Hrecursion4}, the contributions of the first and second dissections are
\begin{multline}
    (-1)^{\frac{|S_0|}{2}+1}  F_{\frac{|S_0|}{2}}(S_0) \qli^{(-)^a}_{|S_{ab}|/2-1}(S_{ab}) \qli^{(-)^b}_{|S_{bc}|/2-1}(S_{bc}) \\
    \qli^{(-)^c}_{|S_{cd}|/2-1}(S_{cd})\prod_{S_{i}\in [0,a]\cup [d,2n-1]}\qli^\pm_{|S_i|/2-1}(S_i)
\end{multline}
and
\begin{equation}
   (-1)^{\frac{|S'_0|}{2}+1}F_{\frac{|S'_0|}{2}}(S'_0) \qli^{(-)^a}_{|S_{ad}|/2-1}(S_{ad})  \prod_{S_{i}\in [0,a]\cup [d,2n-1]} \qli^\pm_{|S_i|/2-1}(S_i)\,,
\end{equation}
where the expression $\prod_{S_{i}\in [0,a]\cup [d,2n-1]}\qli^\pm_{|S_i|/2-1}(S_i)$ represents the contribution of the even-gons between $0$ and $a$, and again between $d$ and $2n{-}1$ (see Figure~\ref{fig: case 1 proof}, left). 
We drop them for notational simplicity since they are identical in the two terms and they do not play a role in the next step. We apply \eqref{eq:DeltaQLi0} and \eqref{eq:coproFphy} to the two terms and compute their contributions to $\Delta_{n-1,1}$ coproduct of the form $\cdots\otimes\qli_1^{(-)^a}(a,b,c,d)$.  The first of these is
\begin{align}
    (-1)^{\frac{|S_0|}{2}+1}F_{\frac{|S'_0|}{2}}(S'_0) \qli^{(-)^a}_{|S_{ab}|/2-1}(S_{ab}) \qli^{(-)^b}_{|S_{bc}|/2-1}(S_{bc}) \qli^{(-)^c}_{|S_{cd}|/2-1}(S_{cd}) \otimes \qli_{1}^{(-)^a}(a,b,c,d)
\end{align}
and the second is exactly the same except the sign factor in front is $(-1)^{\frac{|S'_0|}{2}+1}$. Since $|S_0| - |S'_0| = 2$, these two terms cancel. 

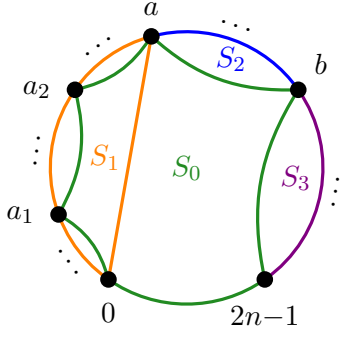
\begin{figure}
    \centering
    \begin{tikzpicture}[scale = 0.45,
    thick,
    dot/.style={circle,fill=black,inner sep=2.2pt},
    ]
    \coordinate (O) at (0,0);
    \def\R{4}

    \coordinate (a)  at ($(O)+(105:\R)$);
    \coordinate (b)  at ($(O)+( 35:\R)$);
    \coordinate (z)  at ($(O)+(235:\R)$);
    \coordinate (n)  at ($(O)+(305:\R)$);
    \coordinate (a1) at ($(O)+(200:\R)$);
    \coordinate (a2) at ($(O)+(145:\R)$);

    \draw[orange, very thick] ($(O)+(235:\R)$) arc (235:105:\R);   
    \draw[blue,   very thick] ($(O)+(105:\R)$) arc (105: 35:\R);   
    \draw[violet, very thick] ($(O)+( 35:\R)$) arc ( 35:-55:\R);   
    \draw[myforestgreen,     very thick] ($(O)+(235:\R)$) arc (235:305:\R);   

    \draw[orange, very thick] (z) -- (a);

    \draw[myforestgreen, very thick] (z)  to[bend right=22] (a1);
    \draw[myforestgreen, very thick] (a1) to[bend right=22] (a2);
    \draw[myforestgreen, very thick] (a2) to[bend right=22] (a);
    \draw[myforestgreen, very thick] (a)  to[bend right=22] (b);
    \draw[myforestgreen, very thick] (b)  to[bend right=22] (n);

    \node[dot] at (a) {};
    \node[dot] at (b) {};
    \node[dot] at (z) {};
    \node[dot] at (n) {};
    \node[dot] at (a1) {};
    \node[dot] at (a2) {};

    \node[above=4pt]   at (a)  {$a$};
    \node[above right=2pt]  at (b)  {$b$};
    \node[below=5pt]        at (z)  {$0$};
    \node[below=6pt]        at (n)  {$2n{-}1$};
    \node[left=5pt]         at (a1) {$a_1$};
    \node[left=5pt]         at (a2) {$a_2$};

    \node[rotate= 127.5] at ($(O)+(217.5:1.10*\R)$) {$\cdots$};
    \node[rotate=  82.5] at ($(O)+(172.5:1.10*\R)$) {$\cdots$};
    \node[rotate=  35  ] at ($(O)+(125  :1.10*\R)$) {$\cdots$};
    \node[rotate= -20  ] at ($(O)+( 70  :1.10*\R)$) {$\cdots$};
    \node[rotate= -100 ] at ($(O)+(-10  :1.10*\R)$) {$\cdots$};

    \node[myforestgreen]      at ( 0.0, 0.0) {$S_0$};
    \node[orange]  at (-2.4,  0.3) {$S_1$};
    \node[blue]    at ( 1.3,  3.2) {$S_2$};
    \node[violet]  at ( 3.2, -0.3) {$S_3$};

    \end{tikzpicture}
    \caption{The dissection illustrating case~(2.3) of~\eqref{eq:Hrecursion4}. We also denote by $S_{1,0} := {\color{myforestgreen} S_0} \cap {\color{orange} S_1}$.}
    \label{fig:case33proof}
\end{figure}

\paragraph{Case (2): rooted.}
We denote the quadrangle by $(0,a,b,2n{-}1)$. This naturally cuts $S$ into three sub-polygons $S_{1},S_{2},S_{3}$. Again, we analyze all potential contributions to the $\Delta_{n-1,1}$ coproduct of the form $\cdots\otimes\qli_1^+(0,a,b,2n{-}1)$.
The three contributions are
\begin{itemize}
    \item[(2.1)] The leading term: $\qli^+_n(P)+\qli^+_{n-1}(P)\log q_P$ in \eqref{eq:Hrecursion4}. Following \eqref{eq:DeltaQLi} and \eqref{eq:DeltaQLi0}, we obtain
    \begin{equation}
    \begin{aligned}\label{eq:coproQLilog}
        & \left(\sum_{ \substack{m_1+m_2+m_3=1 \\ 0\leq m_i} }T^{+,m_1,m_2,m_3}_{0,a,b,2n-1} + T^{+,0,0,0}_{0,a,b,2n-1}\log q_P \right)\otimes \qli_1^+(0,a,b,2n-1)\\
        =&\, \bigg\lbrack 
        {\color{orange} \left( \qli^+_{|S_{1}|/2}(S_{1}) + \qli^+_{|S_{1}|/2-1}(S_{1}) \log q_{S_1} \right) } \qli^-_{|S_{2}|/2-1}(S_{2}) \qli^+_{|S_{3}|/2-1}(S_{3}) \\
        & \qquad + {\color{blue} \left( \qli^-_{|S_{2}|/2}(S_{2}) - \qli^-_{|S_{2}|/2-1}(S_{2}) \log q_{S_2} \right) } \qli^+_{|S_{1}|/2-1}(S_{1}) \qli^+_{|S_{3}|/2-1}(S_{3}) \\
        & \qquad + {\color{violet} \left( \qli^+_{|S_{3}|/2}(S_{3}) + \qli^+_{|S_{3}|/2-1}(S_{3}) \log q_{S_3} \right) } \qli^+_{|S_{1}|/2-1}(S_{1}) \qli^-_{|S_{2}|/2-1}(S_{2})\\
        & \qquad + {\color{red} \qli^+_{|S_{1}|/2-1}(S_{1})\qli^-_{|S_{2}|/2-1}(S_{2})\qli^+_{|S_{3}|/2-1}(S_{3})\log q_{0,a,b,2n-1} } \bigg\rbrack \\
        & \qquad\qquad\qquad\qquad\qquad\qquad\qquad\qquad\qquad\qquad\qquad\qquad \otimes \qli_1^+(0,a,b,2n{-}1) \,,\\
    \end{aligned}
    \end{equation}
    where we have used $q_{P} = q_{S_1} q_{S_3} q_{0,a,b,2n-1} / q_{S_2}$.

    \item[(2.2)] The dissection with $S_0=(0,a,b,2n{-}1)$, giving the $\Delta_{n-1,1}$ coproduct
    \begin{multline}\label{eq:copS0}
        {\color{red} - \log q_{0,a,b,2n-1} \qli^{+}_{|S_1|/2-1}(S_1) \qli^{-}_{|S_2|/2-1}(S_2) \qli^{+}_{|S_3|/2-1}(S_3) }\\ \otimes \qli_1^+(0,a,b,2n{-}1)\,.
    \end{multline}

    \item[(2.3)] The dissection which places $(0,a,b,2n{-}1)$ on the boundary of the root polygon $S_0$. We denote by $(S_1,S_2,S_3)$ the sub-polygons obtained by removing $(0,a,b,2n{-}1)$ from the polygon $S$, as shown in Figure~\ref{fig:case33proof}. The contribution is
    \begin{equation}
    \begin{aligned}
        &\Bigg[\sum_{S_{1,0} \sqcup \cdots \sqcup S_{1,k}= S_1} (-1)^{2+\frac{|S_{1,0}|}{2}} F_{\frac{|S_{1,0}|}{2}+1}\left( S_{1,0}\cup(0,a,b,2n{-}1)\, \right) \prod_{i} \qli^{(-)^{S_{1,i}(1)}}_{|S_{1,i}|/2-1}(S_{1,i})\\  
        & \qquad \times \qli^-_{|S_2|/2-1}(S_2)\,\qli^+_{|S_3|/2-1}(S_3)\Bigg]+ (S_1 \leftrightarrow S_2) + (S_1 \leftrightarrow S_3)\,.
    \end{aligned}
    \end{equation}
    We collect the coproduct of such form
    \begin{equation}
    \resizebox{0.935\textwidth}{!}{$
    \begin{aligned}\label{eq:copS1S2S3}
        &\bigg[-\hspace{-0.4cm}\sum_{S_{1,0} \sqcup \cdots \sqcup S_{1,k}= S_1} \hspace{-0.8cm} {\color{orange} (-1)^{\frac{|S_{1,0}|}{2}} F_{\frac{|S_{1,0}|}{2}}\left(S_{1,0}\right) \prod_i \qli^{(-)^{S_{1,i}(1)}}_{|S_{1,i}|/2-1}(S_{1,i}) } \qli^-_{|S_2|/2-1}(S_2)\,\qli^+_{|S_3|/2-1}(S_3)\\
        &+\hspace{-0.4cm}\sum_{S_{2,0} \sqcup \cdots \sqcup S_{2,k}= S_2} \hspace{-0.8cm} {\color{blue} (-1)^{\frac{|S_{2,0}|}{2}} F_{\frac{|S_{2,0}|}{2}}\left(S_{2,0}^*\right) \prod_i \qli^{(-)^{S_{2,i}(1)}}_{|S_{2,i}|/2-1}(S_{2,i})} \qli^+_{|S_1|/2-1}(S_1)\,\qli^+_{|S_3|/2-1}(S_3)\\
        &-\hspace{-0.4cm}\sum_{S_{3,0} \sqcup \cdots \sqcup S_{3,k}= S_3} \hspace{-0.8cm} {\color{violet} (-1)^{\frac{|S_{3,0}|}{2}} F_{\frac{|S_{3,0}|}{2}}\left(S_{3,0}\right) \prod_i \qli^{(-)^{S_{3,i}(1)}}_{|S_{3,i}|/2-1}(S_{3,i}) } \qli^+_{|S_1|/2-1}(S_1)\,\qli^-_{|S_2|/2-1}(S_2)\bigg]\\
        & \qquad\qquad\qquad\qquad\qquad\qquad\qquad\qquad\qquad\qquad\qquad\qquad\otimes \qli_1^+(0,a,b,2n{-}1)\,,
    \end{aligned}
    $}
    \end{equation}
where $k\geq 0$. The asterisk in $F_{\frac{|S_{2,0}|}{2}}(S^*_{2,0})$ denotes a cyclic shift in the arguments, i.e.~$F_{\frac{|S_{2,0}|}{2}}(S_{2,0}(|S_{2,0}|),S_{2,0}(1),\ldots,S_{2,0}(|S_{2,0}|{-}1))$. Combining equations \eqref{eq:coproQLilog}, \eqref{eq:copS0}, and \eqref{eq:copS1S2S3}, one can see that the {\color{red} red} terms exactly cancel, while the {\color{orange} orange}, {\color{blue} blue}, and {\color{violet} violet} terms combine into zero by applying the relation~\eqref{eq:Hrecursion3} to the polygons {\color{orange} $S_1$}, {\color{blue} $S_2$}, and {\color{violet} $S_3$} respectively.

\end{itemize}

Let us remark that the cancellation of the {\color{blue} blue} terms relies on applying the identity
\begin{equation}\label{eq:Fidentity}
    F_p(2p,1,\ldots,2p{-}1) = - F_p(0,1,\ldots,2p{-}1) \big|_{z_i \to  z_{i+1}, \;\qli^{\pm}\to\qli^{\mp}, \;\log\to-\log}
\end{equation}
to all of the cyclically shifted functions $F_{\frac{|S_{2,0}|}{2}}(S^*_{2,0})$ in~\eqref{eq:copS1S2S3}.  This identity is not obvious but can be proven inductively.  Schematically, one repeats every step in Section~\ref{sec:proof} and this appendix, applying the substitution $z_{i}\to z_{i+1}$ together with $\qli^{\pm}\to\qli^{\mp}$ and $\log\to-\log$ to every $\qli$ and $\log$ function that appears, in order to compute the coproduct and soft limit of the right-hand side of~\eqref{eq:Fidentity}. The result matches the left-hand side as expected.  This proof should be understood to run in parallel with the inductive proof of~\eqref{eq:Hrecursion3}: at step ``$n$'' in the induction, one assumes that~\eqref{eq:Fidentity} is valid for all values $p<n$ to ultimately conclude that it is valid for $n$ as well.

In conclusion, we have confirmed~\eqref{eq:Hrecursion4}, which means that~\eqref{eq:Hrecursion3} is correct up to a single additive constant term.

\subsection{Soft Limit and the Constant Term} 
\label{sec:A2}

The possibility of a missing additive constant in~\eqref{eq:Hrecursion3} can be probed by checking the values of the two sides in a soft limit.  The soft limit $Y_{m,m+1} \to 0$ is equivalent to the points $z_{2m}$ and $z_{2m+1}$  in~\eqref{eq:Cpn} approaching each other, and the soft theorem requires $\psi_n$ to vanish in this limit~\cite{Arkani-Hamed:2023kig}. Here we prove, by induction on $n$, that $F_n(0,\cdots,2n{-}1)$, and hence $\psi_n$, vanishes as $z_{2m} - z_{2m+1} \to 0$ for any $1\leq m\leq n-1$. 

We start from the second term on the right-hand side of~\eqref{eq:Hrecursion3}. For an arbitrary dissection $\{S_0,\ldots,S_k\}$ of the polygon $(0,\ldots,2n{-}1)$ into even sub-polygons, as listed in~\eqref{eq:Hrecursion3}, exactly one sub-polygon $S_j$ (with $0\leq j\leq k$) contains both $z_{2m}$ and $z_{2m+1}$. We treat the cases $j=0$ and $j>0$ separately:
\begin{itemize}
    \item When $j = 0$, by the inductive hypothesis $F_{\frac{|S_0|}{2}}(S_0)$ vanishes in the soft limit, and hence so does the whole term. 
    \item When $j > 0$, we examine $\qli^{(-)^{S_j(1)}}_{|S_j|/2-1}(S_j)$ as $z_{2m} - z_{2m+1} \to 0$. Following Rudenko's arborification algorithm, $z_{2m}-z_{2m+1}$ enters only in the numerators of the weighted-symbol arguments. Hence each term in the $\Li_k(W)$ expansion has at least one argument that vanishes; from the series expansion of the multiple polylogarithm in~\eqref{eq: MPL_series_expansion}, it is clear that the multiple polylogarithm vanishes whenever any of its arguments does. Therefore $\qli^{(-)^{S_j(1)}}_{|S_j|/2-1}(S_j)$, and hence the whole term, vanishes in the soft limit.
\end{itemize}

For the first term on the right-hand side of~\eqref{eq:Hrecursion3} the argument is similar: both $\qli^+_{n}(0,\ldots,2n{-}1)$ and $\qli^+_{n-1}(0,\ldots,2n{-}1)$ vanish in the soft limit, since each term in their expansion in terms of multiple polylogarithms has at least one argument going to zero; the apparent $\log q_{0,\ldots,2n-1}$ divergence multiplying $\qli^+_{n-1}$ is suppressed by these zeros. Finally, we can easily check the initial seed \eqref{eq:Ffour} of the recursion vanishes by direct computation. We conclude that the right-hand side of~\eqref{eq:Hrecursion3} vanishes in every soft limit, fixing the constant term that is otherwise invisible to the coproduct.

\bibliographystyle{JHEP}

\bibliography{reference}

\end{document}